\documentclass[12pt]{article} 
\usepackage[utf8]{inputenc}
\usepackage{amsmath,amssymb,amsfonts} 
\usepackage[UKenglish]{babel} 
\usepackage[T1]{fontenc}
\usepackage{setspace}
\usepackage{xcolor}
\usepackage{graphicx}
\usepackage{lmodern}
\usepackage{booktabs}
\usepackage{tabularx}
\usepackage{caption}
\usepackage{hyperref}
\usepackage[T1]{fontenc}
\usepackage{ae,aecompl}
\usepackage{subfigure}
\usepackage{fullpage}
\usepackage[toc,page]{appendix}
\usepackage{epstopdf}
\usepackage{import} 

\usepackage{todonotes}

\newcommand{\Stem}{\mathfrak R} 
\newcommand{\final}{{\mathcal N}_f}
\newcommand{\initial}{{\mathcal N}_i}
\newcommand{\infin}{{\mathcal N}_{i,f}}
\allowdisplaybreaks 

\graphicspath{{figures/}}

\newcommand{\av}[1]{\langle #1 \rangle}

\newcommand{\Sa}{\mathcal{S}}
\newcommand{\anti}{\mathcal{A}}

\newcommand{\twod}{{\mathcal S}_{\mathrm {2d}}} 
\newcommand{\avS}{\langle \twod \rangle}
\newcommand{\partn}{{\mathcal Z}}

\newcommand{\tO}{\Omega_{\mathrm{cov}}}
\newcommand{\mA}{\mathfrak A} 

\newcommand{\tm}{\widetilde m}
\DeclareMathAlphabet{\mathpzc}{OT1}{pzc}{m}{it}

\newcommand{\mc}[1]{\mathcal #1}

\title{The  Hartle-Hawking wave function \\  in 2d causal set quantum gravity} 
\author{Lisa Glaser$^{a}$ and Sumati Surya$^{b}$\\ 
$^a$ School of Mathematical Sciences
University of Nottingham, UK \\ \& Niels Bohr Institute, Copenhagen, Denmark \\
$^b$ Raman Research Institute, Bangalore, India}
\begin{document}
\parskip 0.3cm

\maketitle

\begin{abstract}
  We define the Hartle-Hawking no-boundary wave function for causal set theory (CST) over the
  discrete analogs of spacelike hypersurfaces.  Using Markov Chain Monte Carlo and numerical
  integration methods we analyse the wave function in non-perturbative 2d CST.  We find that in the
  low temperature regime it is dominated by causal sets which have no continuum counterparts but
  possess physically interesting geometric properties. Not only do they exhibit a rapid spatial
  expansion with respect to the discrete proper time but also a high degree of spatial
  homogeneity. The latter is due to the extensive overlap of the causal pasts of the elements in the
  final discrete hypersurface and corresponds to high graph connectivity.
  Our results thus suggest new possibilities for the role of quantum gravity in the observable
  universe.
\end{abstract}

\section{Introduction}

The Hartle-Hawking (HH) prescription for the ground state wave function over closed 3-geometries
$(\Sigma, h)$ is the Euclidean functional integral over \mbox{4-geometries} $(M,g)$
\begin{equation} 
\Psi_0(h_{ab},\Sigma)= A \sum_{M} \int dg^E e^{-I_E(g)}
\label{hhcont} 
\end{equation} 
where $\partial M=\Sigma, g|_\Sigma=h$, $I_E(g)$ is the Euclidean Einstein action and $A$ is a
normalisation constant \cite{hh}.  $\Psi_0(h_{ab}, \Sigma)$ is thus a functional over all closed
3-geometries and is the initial state of the universe from which further evolution of the wave
function can be (uniquely) determined.  This ``no-boundary'' proposal thus does away with
ambiguities coming from boundary conditions, since there is only one ``final'' boundary $\partial M$
in (\ref{hhcont}), and no ``initial'' boundary. In analogy with quantum field theory, this proposal
uses the Euclidean path integral for defining the ground state. Although this implies an ambiguity
in assigning a time to the boundary, the future evolution from this state is expected to be
Lorentzian.

While the simplicity and ingenuity of this proposal is undeniable, the continuum path integral in
notoriously ambiguous and needs to be regulated.  Different discrete approaches to quantum gravity 
like simplicial quantum gravity, causal dynamical triangulations and causal set theory(CST) choose
different ``regularisation schemes'' \cite{david,lollambjorn,blms}. In particular, CST  posits a fundamental
discreteness where the spacetime continuum is replaced by a locally finite partially ordered set or
causal set (\textit{causet} for short) and the path integral, by a sum over causets
\cite{blms,reviewone,reviewtwo}.  This is the setting in which we will examine the fully non-perturbative contributions to the HH wave function.
 
It is important to emphasise here that CST differs from other discrete approaches in some critical
ways. Causality plays an important role: in a causet which is approximated by a continuum spacetime
two elements are \textit{related} if there is a causal relation between them, otherwise not. Thus
there are no ``spacelike'' nearest neighbours.  Key aspects of the CST approach that are useful to
keep in mind: (i) a \textit{fundamental} (Lorentz invariant \cite{li}) discreteness with the
continuum arising as an approximation via a Poisson process
(ii)  the sum over causets includes those with no continuum counterpart and continuum geometries differing
on scales smaller than the cut-off correspond to the same causet  (iii) there is no way to
``Euclideanise'' a causet -- it is fundamentally Lorentzian, and (iv) unlike the fixed valency dual
graphs of fixed-dimension triangulations, causets are graphs with varying valency (not necessarily
even finite).  All these features make CST distinct from other discrete approaches to quantum
gravity.

In order to extend the HH prescription to CST, the continuum path integral has to first be replaced
by a sum over causets, which must satisfy the analog of the no boundary condition. Note that
since there are no Euclidean causets, the HH prescription can only be implemented over
Lorentzian structures, though the dynamics can be Euclideanised.  We thus need to define a spatial
boundary in a causet $C$ in analogy with $\partial M$. An {\sl antichain} or subset of unrelated
elements in $C$ is a spacelike hypersurface. For it to be a past or future boundary, it must further
be {\sl inextendible} in that one cannot add more elements to it\footnote{Adding more elements to an
 this  antichain would not allow it to remain an antichain. In the poset literature the term  ``maximal'' is also used.}, and such that no element lies to {\it its} past or future,
respectively \cite{antichain}.  A final spatial boundary of a spacetime region is thus represented
in $C$ by a future-most inextendable antichain $\anti_f$ in $C$ and an initial spatial boundary by a
past-most inextendable antichain $\anti_i$.  While an antichain by itself appears to contain scant
information, being intrinsically defined only by its cardinality, its connectivity to the bulk elements in
$C$ contain significant geometric information. To implement the no boundary proposal in CST we
restrict the sum over causets to \textit{ no-boundary causets}, i.e., those with
$\initial=|\anti_i|=1$ ({\sl originary causets}) and a fixed $\final=|\anti_f|$.  That
$\initial=|\anti_i|=1$ could be compatible with the no-boundary condition might seem at first sight
counterintuitive, but it is the only natural choice in the discrete setting of CST.  Indeed, it is
the exact discrete analog of what the authors of \cite{hh} refer to as an initial spatial ``zero''
geometry, a single point, which captures the idea of a universe emerging from nothing.  Any other
choice for $\initial$ would violate this requirement.

In the continuum the Euclidean path integral serves two purposes: (i) the no-boundary topology does
not support a singularity free causal Lorentzian geometry \cite{geroch,rds-topology}, and (ii)
Euclideanisation yields a probability measure.  As we have stated above, there is no analog of a
Euclidean causet since  every causet is both causal (thence Lorentzian) and non-singular.  Thus, the
sum over causets remains Lorentzian. However,
the quantum measure  $\exp(i S(C)/\hbar)$, where $S(C)$ is the action of the causet  can be made
into a probability measure  by Wick rotating a 
supplementary 
variable $\beta$ which multiplies $S(C)$ and plays the role of the inverse  temperature
\cite{sorkinlouko,2dqg}. Using this, we   define the HH wave function in CST as
\begin{equation}
\Psi^{(N)}_0(\final, \beta) \equiv A \sum_{C\in \Omega_N}  e^{-\frac{1}{\hbar}\beta S(C)}  
\label{hh}
\end{equation} 
where $\Omega_{N}$ is the space of $N$ element no-boundary causets, $S(C)$ is a causet action and
$N$ is fixed as in unimodular gravity.  In the continuum this corresponds to keeping the spacetime
volume $V$ fixed as one does in unimodular gravity. $N$ thus acts as a proxy ``time'' label. What
the meaning of such a time label is, and whether it can be given a covariant interpretation are
questions we will attend to at the end of this paper.  

For now we note that $\Psi^{(N)}_0(\final, 0)$ is the (normalised) uniform distribution over
$\Omega_N$. When there is no restriction on $\initial$ or $\final$ this distribution is dominated in
the asymptotic limit by the Kleitman-Rothschild posets \cite{kr}, but the behaviour when $\final$
and $\initial$ are fixed is not known.  Note also that while the introduction
of $\beta$ appears at first ad-hoc as opposed to the straightforward choice of $\beta=1$, it is
known to play a non-trivial role in the scaling behaviour of 2d quantum gravity \cite{2drg}; an RG
analysis suggests that it possesses a fixed point that differs from $\beta=1$. Carrying out  such an
analysis in the present context of the HH wave function should yield similar non-trivial
behaviour, but is outside the scope of this present work.

In this work we evaluate the HH wavefunction in 2d CST where the causets are restricted to the set
of \textit{2d orders} $\Omega_{2d}$. This theory has proved to be a non-trivial testing ground for
CST \cite{2dqg,2dorder}.  It includes causets that are approximated by continuum 2d spacetimes
in an open disc as well as those that have no continuum counterpart.  An $n$d order is obtained by
intersecting $n$ total orders and it is indeed a coincidence that this order theoretic dimension
coincides with manifold dimension for $n=2$ in the sense described above. An interesting feature of
2d orders is that the uniform distribution ($\beta=0$) is dominated by {\sl 2d random orders} which
are causets that are approximated by 2d flat spacetime \cite{2dorder}.  In \cite{2dqg} it was
shown that as $\beta$ increases from zero the 2d random orders dominate until a critical value of
$\beta$ at which point  there is a phase transition from this continuum phase to
one that is distinctly non-continuum like. Thus,  varying  $\beta$ can have a strong influence on the
dominant contributions to the partition function, a feature that is echoed in this current work.  

As we will show  it is possible to calculate $\Psi^{(N)}_0(\final, \beta)$ analytically in 2d CST for the largest 
values of $\final$. However, the calculation becomes rapidly more difficult for smaller values of
$\final$, and we must resort to numerical methods.  We use MCMC methods for the equilibrium
partition function to obtain the expectation value of the 2d action $\avS_\beta(\final)$ as a
function of $\final$ and $\beta$ for a fixed causet size $N=50$.  We then evaluate the 2d
partition function at $\beta=0$, $\partn_0(\final)$ by counting the number of 2d random orders with
a given $\final$ for a sufficiently large ensemble.  From this we can then calculate the HH wave
function for 2d gravity
\begin{equation} 
\Psi^{(N)}_0(\final,\beta) = A \partn_\beta(\final) =
A \partn_0(\final) \exp^{-\int_0^\beta d\beta' \avS_\beta'(\final)}
\end{equation} 
by performing the requisite numerical integration.  While the MCMC methods thermalise well for
most values of  $\final$ this is not so  for the largest values of $\final$ for which we use 
our analytic results.  

In \cite{2dqg}, it was shown that $\avS_\beta$ with no constraint on $\final$ exhibits a phase
transition. Here we find that the same is true when fixing $\final$ to values sufficiently smaller
than $N$. However, there is a new non-trivial dependence of the phase transition temperature with
$\final$ with the former achieving a minimum (critical) value $\beta_c$ at some intermediate value
of ${\final}_c$.  This feature gives rise to a surprisingly well defined peak in
$\Psi^{(N)}_0(\final,\beta)$ whose center shifts as one approaches $\beta_c$ from a small $\final$
value to one close to ${\final}_c$. These two peaks lie at causets that correspond to random 2d
orders (approximated by open regions of 2d Minkowski  spacetime) and those that have no continuum
counterpart, respectively. It is indeed the detailed character of the latter that is particularly interesting. Despite being non-manifold
like, these causets share important features of the early universe. Not only do they exhibit a
rapid spatial growth for very small proper time, but their pasts are highly overlapping, suggesting
a far greater degree of causal connectivity than could be obtained in the continuum. This results in
a high degree of spatial homogeneity without recourse to a  {\it continuum} inflationary scenario.

Although our calculations are restricted to a 2d universe and any generalisation to include higher
dimensions will undoubtedly be very non-trivial, our results nevertheless are highly
suggestive. They allow  a clear physical interpretation and provide a novel insight into the possible
role of quantum gravity in the early universe.  Thus, while the calculation is one of a
``proof of principle'', the results themselves are far more suggestive than one might have expected.

In a causet it is not only the number of elements $N$ that decide on its complexity, but also the
number of possible relations $N(N-1)/2$. For $N=50$ this number is  $1225$, which is still
insignificant  in cosmic terms, but puts a practical limit on the computation. This is
especially true of the HH wave function which depends on several parameters, all of which must be
explored. Judging from earlier work on 2d CST where one obtains very good scaling behaviour with
$N$, one would guess that the qualitative  features of our results will remain the same
\cite{2drg}.  Of course, a further computation with larger values of $N$ should give us more confidence in
these results. 

Finally, is important to ask whether the HH wave function thus defined  has a truly covariant
interpretation, in particular one that survives the subsequent evolution of the causet.  For example
(as in the sequential growth models of  \cite{csg}) it is possible for a 
causet with final boundary $\anti_f$ to evolve to contain elements that are not causally related to
those in $\anti_f$.  Hence the latter is no
longer an ``initial condition''  or the ``summary of the past'' as a Cauchy hypersurface should be.  In
order to ensure such covariance the HH wave function must give the measure over all countable
causets (finite to the past and countably infinite to the future) in which $\anti_f$ is the
appropriate discrete analogue of a Cauchy hypersurface.  We find that this possible to achieve, at
least partially, using formulations of measure theory on causets 
\cite{observables,antichain}.

In Section \ref{two} we lay down the basics of 2d CST. In Section \ref{three} we show the analytic
calculation of the wave function  for general $N$ for $\final=N-1,N-2,N-3$.   We then describe the MCMC
calculations, and the numerical integration methods required. We present our main results and conclusions
in Section \ref{four}.  We end in Section \ref{five} with a measure theoretic interpretation of the  HH proposal and
how we can view its as the quantum measure of a covariant event \cite{qmeasure}.   

\section{2d Causet Theory} 
\label{two} 

A causet $C$ is a locally finite partially ordered set. This means that $C$ is a set  with a
relation $\prec$ which is,   for any $x,y,z \in C$,  
\begin{enumerate}  
\item Reflexive: $x \prec x$. \label{ref}  
\item Acylic: $x \prec y$ and $y \prec x $ implies that $y=x$. \label{acy}       
\item Transitive: $x\prec y$ and $y \prec z$ implies $x \prec z$.  \label{tran}   
\item Locally finite: If $\mathrm{Fut}(x)\equiv \{y| x\prec y \} $ and $\mathrm{Past}(x)\equiv \{y|
  y\prec x \} $, the set $I(x,y)=\mathrm{Fut}(x) \cap \mathrm{Past}(y) $ is of finite cardinality.  
\end{enumerate} 
It will be useful in what follows to refer to the set $I(x,y)$ as an {\sl interval}, in analogy with
the Alexandrov interval in the continuum. The condition of local finiteness thus ensures a
fundamental spacetime discreteness, so that a spacetime interval of finite volume contains a finite
number of spacetime atoms.

A causet $C$ is said to be approximated by a spacetime $(M,g)$ at fundamental scale $V_c$ if it
admits a ``faithful embedding''.  By this we mean that $C$ should be generated from $(M,g)$ via a
Poisson process (or sprinkling) where the order relation between the elements is induced by the
causal order in $(M,g)$.  Thus, the probability of there being $n$ elements in a spacetime volume
$V$ is $P_V(n) = (1/n!) \exp^{-V/V_c} (V/V_c)^n$, for a given cut-off $V_c$, so that $\langle n
\rangle = V/V_c$, i.e., there is a correspondence between the average number of elements to the
volume of the spacetime region. This summarises the discreteness hypothesis: finite volumes in the
continuum contain only a finite number of fundamental spacetime atoms.  The choice of the Poisson
distribution means that discreteness also retains Lorentz invariance in a fundamental way \cite{li}.

In (\ref{hh}) $\Omega_N$ is the set of all $N$-element causets which have a single initial element,
i.e., $\initial=1$ and some fixed $\final$ final elements. Without further restrictions on
$\Omega_N$ there is no specification of continuum dimension which is instead expected to be
emergent.  In this work we will focus instead on a simplification to 2d CST in which $\Omega_N$ is
further restricted to the set of {\sl 2d orders} $\Omega_{2d} \subset \Omega_N$ \cite{2dqg,2dorder}.
These causets are dimensionally and topologically constrained in the continuum approximation to an
open disc in 2-dimensional Minkowski spacetime. While $\Omega_{2d}$ contains causets that have a
continuum approximation, they also contain those that have no continuum correspondence. These will
in fact play a crucial role in the results of Section \ref{four}.  A 2d order is defined as follows.
Let $S=\{1, \ldots, N\}$ be a base set. For $u_i, v_i \in S$, $U=(u_1, \ldots u_N)$ and $ V=(v_1,
\ldots v_N)$ are totally ordered w.r.t. the natural ordering $<$ in $S$, i.e., for every $u_i,u_j
\in U$, either $u_i\!<\!u_j$ or $u_j\!<\!u_i$ and similarly for $V$.  A \textit{2d order} $C = U
\cap V$ is a causet with elements $e_i=(u_i,v_i)$ such that $e_i \prec e_j$ in $C$ iff $u_i < u_j$
{\it and} $v_i < v_j$. It is obvious from this presentation of the 2d order, and taking $u_i,v_i$ to
be light cone coordinates, that every 2d order admits an embedding into 2d Minkowski
spacetime. However, this need not be a faithful embedding, so that not all 2d orders admit a
continuum approximation. An important example of one that does admit a continuum approximation is a
2d random order, with the $u_i$ and $v_i$ chosen at random and independently from $S$. This is
approximated by an Alexandrov interval in 2d Minkowski spacetime \cite{2dorder}.

The 2d causet version of the discrete Einstein-Hilbert action for a causet $C$ \cite{bd,dg} is  
\begin{equation} 
\frac{1}{\hbar}\twod(N,\epsilon) = 2
\epsilon \biggl(N-2 \epsilon \sum_{i=1}^N f(i,\epsilon) N_i\biggr), 
\label{action} 
\end{equation}  
where $N_i$ is the number of $(i-1)$-element interval in $C$, $\epsilon ={l_p}/{l} \in (0,1]$, with
$l_p$ the Planck scale and $l\! >\! l_p$ the non-locality scale and
\begin{equation} 
f(i+1,\epsilon)=
 (1-\epsilon)^i\biggl(1-\frac{2i\epsilon}{(1-\epsilon)}+
 \frac{i(i-1)\epsilon^2}{2(1-\epsilon)^2}\biggr). 
\end{equation} 
The non-locality scale $l$ is that at which the locality of the continuum should arise, and below
which the causet has non-manifold like properties.  

As discussed in \cite{2dorder,2dqg} the uniform distribution over $\Omega_{2d}$ (i.e., $\beta=0$)
without the no-boundary condition is dominated by random 2d orders, which means that flat spacetime
dominates in this limit. However, as shown in \cite{2dqg} as $\beta$ increases, there is a phase
transition at $\beta_c$ and one shifts from a  continuum phase to a fundamentally discrete phase which
possesses regular, or ``crystalline'' structure.  As we will see, this fundamental feature
remains when the no-boundary condition is imposed. However, the transition temperature shows a
non-trivial dependence on $\final$, which is key to the results that we will find.

\section{Calculating the Hartle Hawking Wavefunction} 
 \label{three}

\subsection{Analytic Calculation} 

The HH wavefunction for 2d CST can  be evaluated analytically for the largest values of $\final$ but
becomes harder for smaller values. This is because the number of ``bulk elements'' ', i.e.,
those not in $\anti_i \cup \anti_f$ becomes larger, which means that one has  many different
2d-orders in the bulk that contribute to the sum.  On the other hand, for the largest values of $\final$
the MCMC simulations become less reliable, since the thermalisation times become significantly
larger. Thus, the analytic calculations become important in supplementing our  numerical
calculations.  

\begin{figure}
\centering{\includegraphics[width=0.4\textwidth]{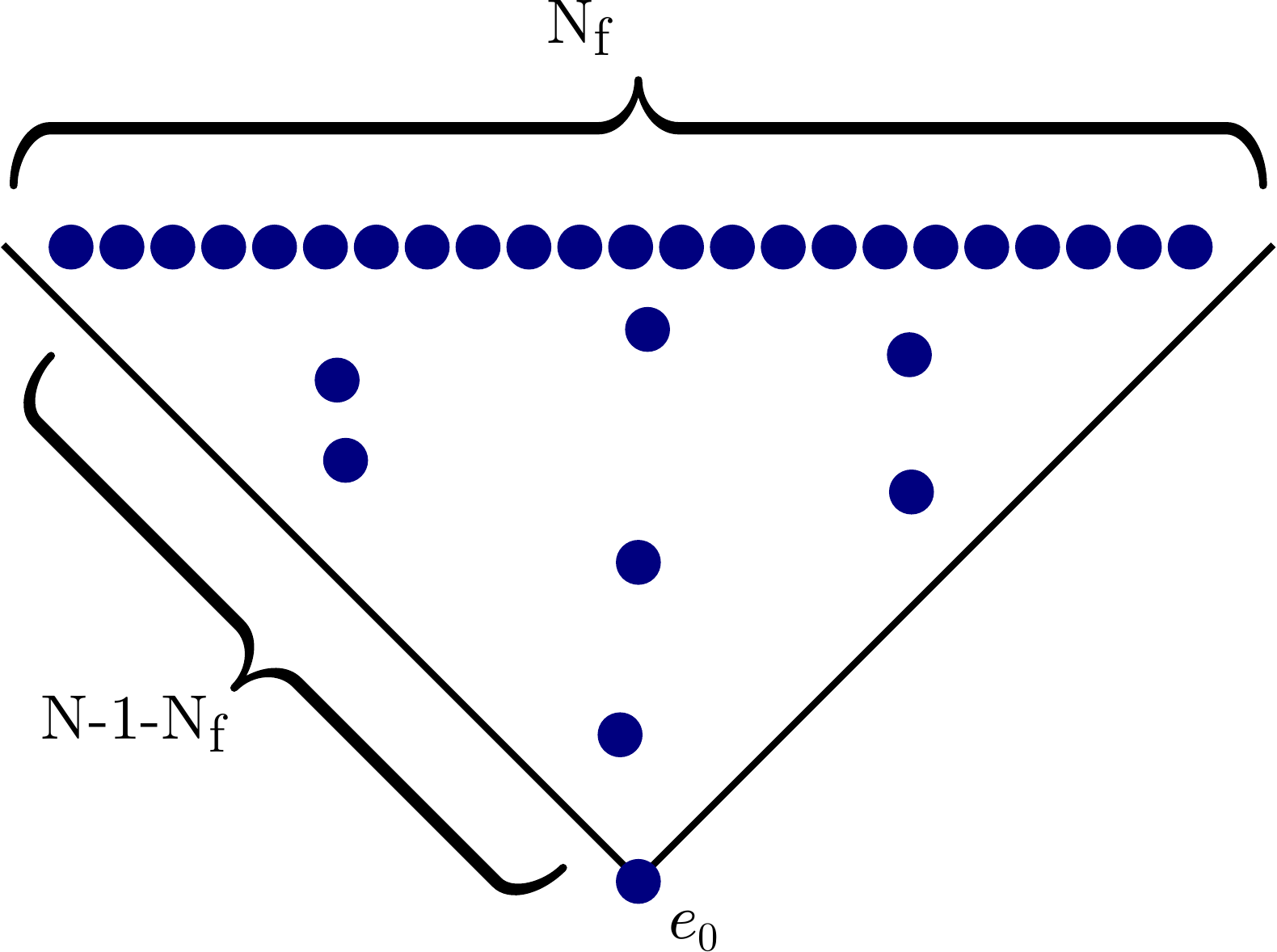}\hspace{10pt}}
\caption{An illustration of a  2D order in  $\Omega_N$ with  $N_i=1$ and arbitrary $\final$ }
\label{fig:Nfgeneral}
\end{figure}

We now calculate the HH wave function for $\final=\!N\!-\!p$, $p=1,2,3$. In what follows the number
of  distinct labeled 2d orders contributes a ``multiplicity'' which we include in the sum. This is a
choice of measure that is most suited to the numerical calculations that follow and was adopted in
\cite{2dqg}. By including this multiplicity in the full measure, it is important to note that there
is no violation of covariance.  Since the element in $\anti_i$ is to the past of all other elements
it is uniquely labelled as  $e_0=(0,0)$ (see   Fig(\ref{fig:Nfgeneral})). The choice of labelings of
the other elements in the causet depends on several factors as we will see below. 

\noindent $\mathbf{p=1:}$  Here since there are no bulk elements, there is only one labelled causal
set. All relabellings of the elements in $\anti_f$ are merely automorphisms.  The only non-vanishing
$N_i$ which contributes to the action (\ref{action}) is the number of links $N_1=N-1$. The action
thus simplifies to
\begin{equation} 
\frac{1}{\hbar}\twod(N,\epsilon) =2 \epsilon N(1- 2\epsilon)+4 \epsilon^2, 
\end{equation} 
so that 
\begin{equation} 
\Psi_0(N-1)=A e^{-\beta R}, 
\label{N1}
\end{equation}   
where $R=2 \epsilon N(1- 2\epsilon)+4 \epsilon^2$.  We have  suppressed the $N$ and $\beta$
dependence in $\Psi_0$ and we will do so from now on. 

\noindent $\mathbf{p=2:}$ Here there is a single bulk element, $e_1=(u_1,v_1)$.
Although it must be to the future of $e_0$ it need not be to the past of every element in
$\anti_f$. Thus there are multiple causets with this boundary condition which will contribute to
$\Psi_0(N-2)$.  These causets can be characterised by the number of elements $l$ to the future of
$e_1$.  If $F$ is this set of future elements of $e_1$ (necessarily in $\anti_f$) then $l \equiv
|F|$, $l \in (1, \ldots, N-2)$.  For a given $l$, the number of inclusive intervals are:
$N_1=N-1$, $N_2=l$ and $N_i=0, \, \forall \, i>2$, which means that the action depends only on $l$
(apart from $N,\epsilon$).  However, $l$ does not determine the causet uniquely, since there is
a non-trivial multiplicity $\mu_l$ associated to the number of distinct labelled 2d orders for a
given $l$.

\begin{figure}
\centering
\includegraphics[width=0.5\textwidth,angle=45]{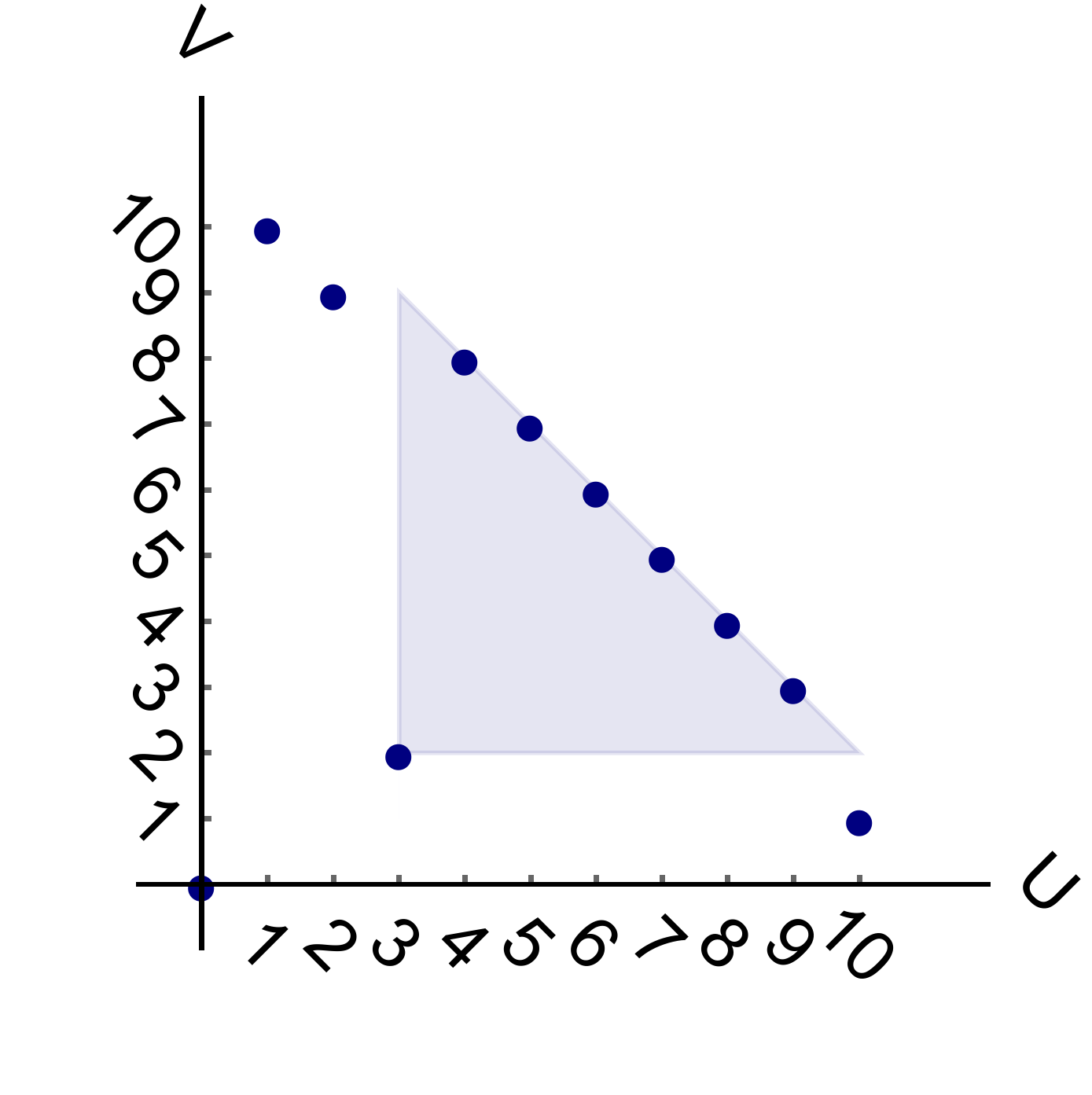}
\caption{\label{fig:N-2case}Illustration of a $N-2$ dimensional configuration. The future of the bulk element is shaded.}
\end{figure}

$\mu_l$ can be obtained as follows.  For every $ e =(u,v) \in F $, $u > u_1 $ and $ v >v_1$ so that
that $u_1,v_1 \in [1, \ldots N-1-l]$. Moreover, for the remaining $N-2-l$ elements in $S=
\anti_f\backslash F$, for $e_s=(u_s,v_s)$,  $u_s<u_1 \Rightarrow v_s>v_1$ and $u_s>u_1 \Rightarrow
v_s<v_1$. There are  $u_1-1+v_1-1$ such  possibilities and these must be equal to the number of
elements $N-2-l=|S|$ since the only other remaining elements are either to the future of $e_1$ or to
its past. This fixes $v_1=N-u_1-l$.  Indeed, there are no further constraints that can be put on
$u_1$: every choice of $u_1$ fixes precisely which elements lie in $F_1$. Wlog, let
us order the set of $u$-values in $F$ such that $u_{i_1} < u_{i_2}< \ldots <u_{i_l}$. Since $F$ lies
in $\anti_f$ this means that  $v_{i_1} > v_{i_2}> \ldots >v_{i_l}$. These values are contiguous: for
any $e =(u,v) \in \anti_f \backslash F$  either $u<u_{i_1}$  or $u>u_{i_l}$ since otherwise if
$u_{i_m} < u< u_{i_{m+1}}$ then $u> u_1$ and  $v> v_{i_{m+1}}>v_1$ which is not possible. For a
fixed $u_1$ this then uniquely fixes the labeling of all other elements. 
That the $u_{i_m}$s are contiguous also becomes obvious from
the fact that every 2d order embeds (albeit non-faithfully) into 2d Minkowski spacetime (see
\mbox{Figure \ref{fig:N-2case}}). 
Hence $\mu_l=(N-1-l)$. The expression for $\Psi_0(N-2)$ is then easily evaluated since it involves 
sums over a geometric series  and their derivatives 
\begin{equation} 
\Psi_0(N-2) = \frac{A e^{-\beta (R-Q)}}{(1-e^{\beta Q})^2} \biggl( N-2-(N-1)
e^{\beta Q}  + e^{\beta Q(N-1)} \biggr)  
\end{equation} 
where  $Q=4 \epsilon^2(1-3\epsilon)$ and $R$ is given as in (\ref{N1}).

\begin{figure}
\subfigure[Antichain with overlapping
futures]{\includegraphics[width=0.23\textwidth,angle=45]{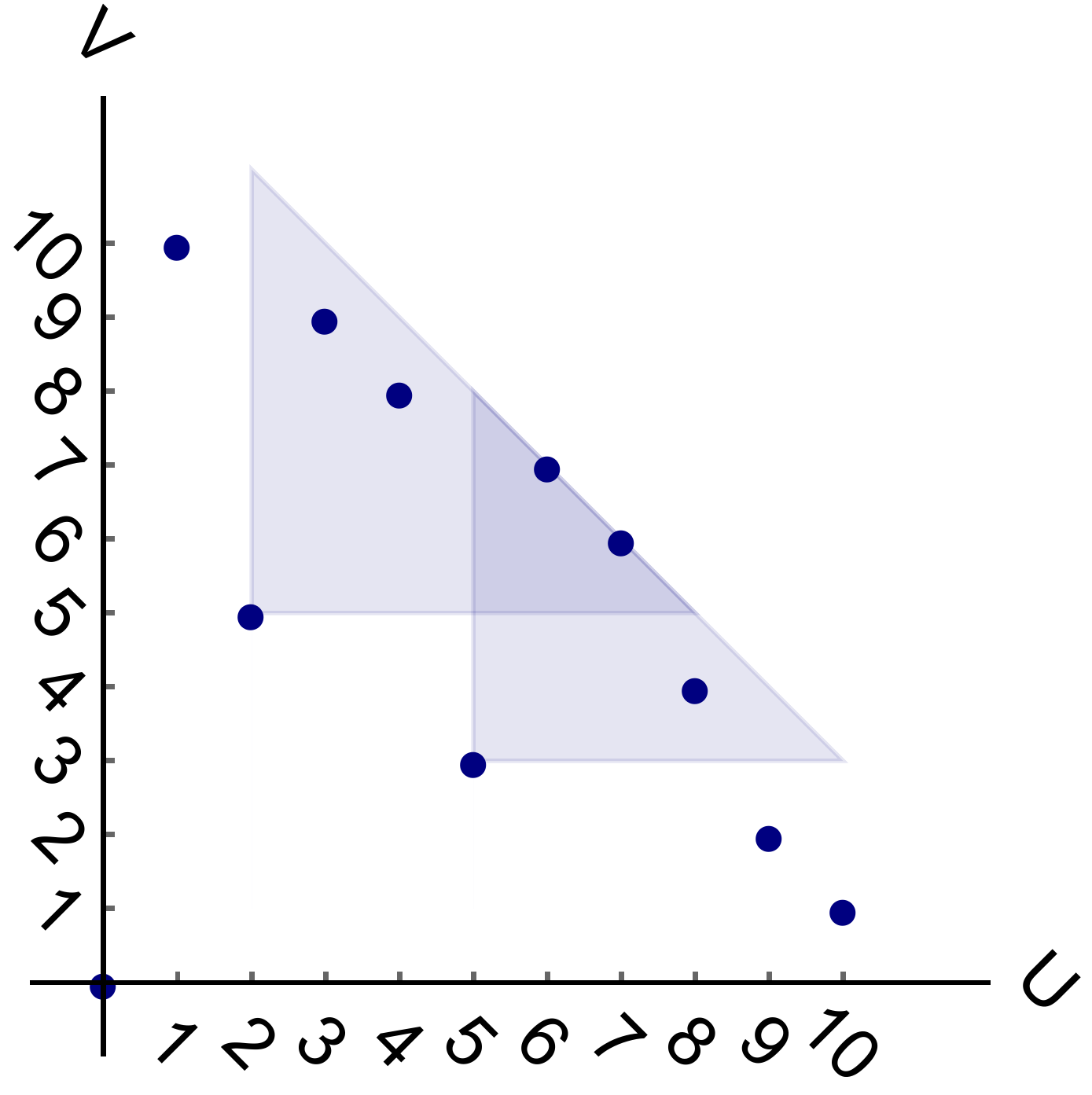}} 
\hskip0.1cm \subfigure[Antichain with non-intersecting futures]{\includegraphics[width=0.23\textwidth,angle=45]{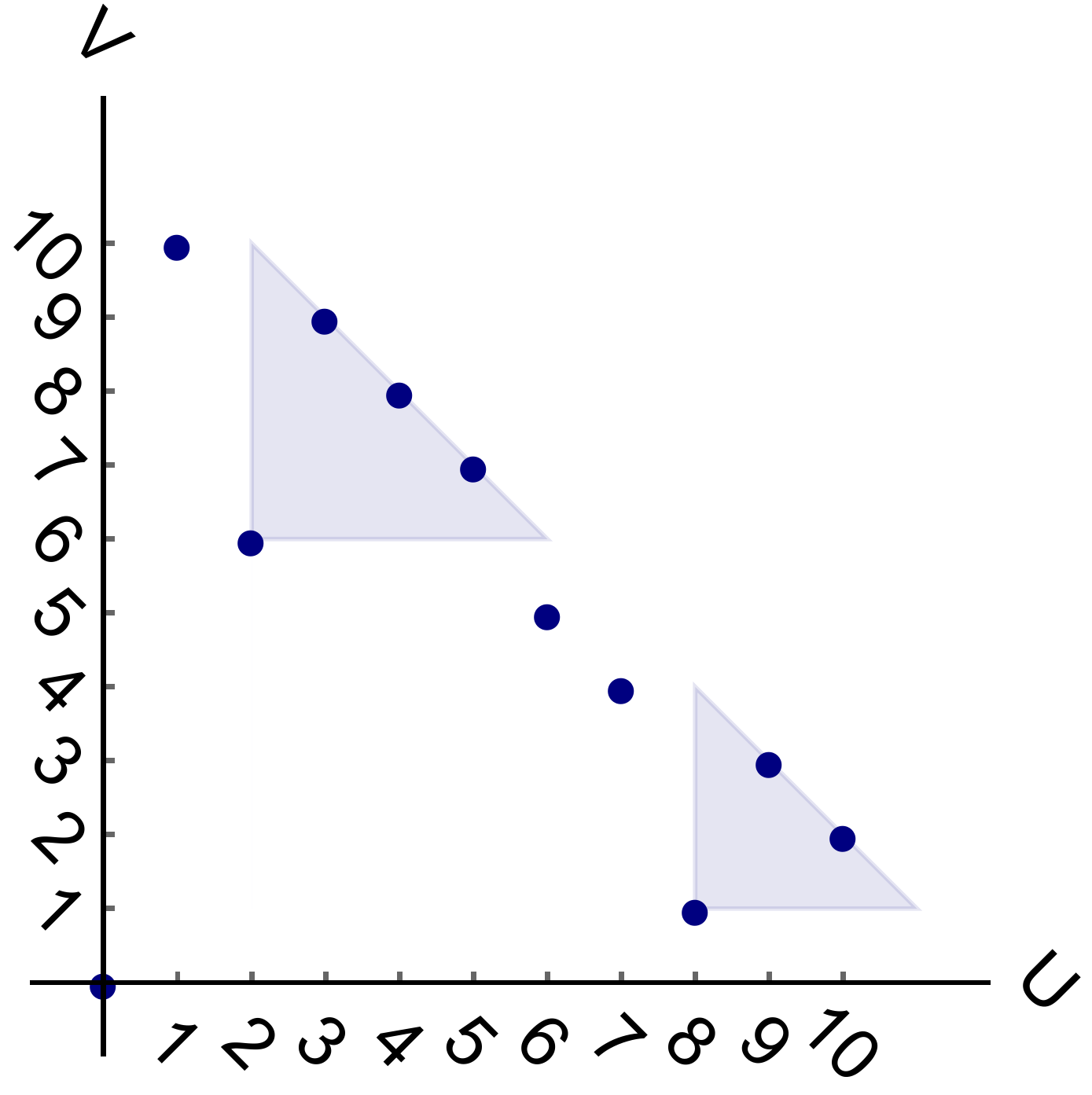}}
\subfigure[Chain]{\includegraphics[width=0.23\textwidth,angle=45]{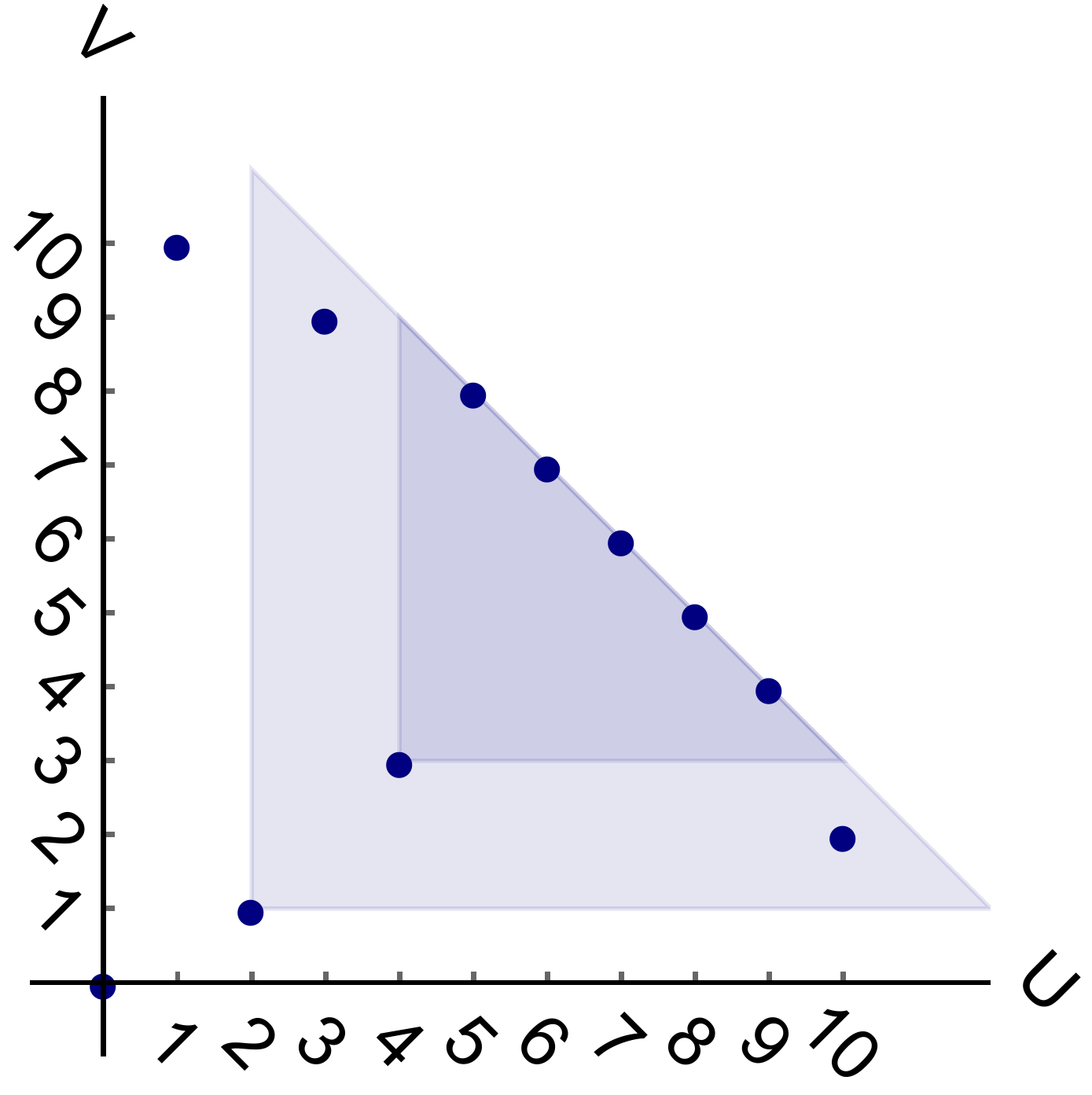}}
\caption{\label{fig:N-3case}Illustration of the three types of configuration possible with $2$ bulk elements. The future of the bulk elements is shaded.}
\end{figure}

\noindent $\mathbf{p=3:}$ This  calculation is more involved. For one, the bulk elements $e_1,e_2$ can
form (a) an antichain or (b) a chain. These possibilities are shown in Figure \ref{fig:N-3case} . In both cases we  will denote the future of the $e_{1,2}$ in
$\anti_f$ as $F_{1,2}$ and
let $l_{1,2}=|F_{1,2}|$.  

For (a) there are two possibilities, (i) $F_1 \cap F_2 \neq 0$ so that $|F_1\cap F_2|=m \leq
max(l_1,l_2)$ and (ii) $F_1 \cap F_2 = 0$ with a ``spacing'' $\tm$ between the maximum $u$
values in $F_1$ and the minimum $u$ value in $F_2$.  This spacing comes from the fact that the
$u$ and $v$ values in $F_{1,2}$ are contiguous as we saw for the $\anti_f=N-2$ case. For (i) the
non-zero abundances of order intervals are $N_1=N-1-m$, $N_2=l_1+l_2-2m$ and $N_3=m$ and for (ii)  
$N_1=N-1$, $N_2=l_1+l_2$ and $N_3=0$. These can be combined to case (i) since in case (ii) $m=0$. 

Next, we calculate the multiplicity $\mu_{l_1,l_2,m,\tm}$.  In both cases (i) and (ii) the
arguments in the $\anti_f=N-2$ case can be used to see that $u_1 \in 
[1, \ldots, N-1-l_1], v_1=N-l_1-u_1$ and   $u_2 \in [1, \ldots, N-1-l_2], v_2=N-l_2-u_2$ independently.  
Assume $u_1 \prec u_2$.  Then the contiguous values of $u$ in $F_1$ and $F_2$ are therefore such
that the minimum value $u_{j_1}^{(1)}$ in $F_1$ is less than or equal to the minimum value
$u_{j_1}^{(2)}$ in $F_2$. If $u_{s_r}^{(1)}$ denote the values of $u$ in $\anti_f\backslash F_1$  with
$r=1,\ldots,|\anti_f\backslash F_1|$ one
has the ordering 
\begin{equation} 
u_{s_1}^{(1)}<u_{s_2}^{(1)}<\ldots u_{s_p}^{(1)}< u_{j_1}^{(1)}<u_{j_2}^{(1)} \ldots
u_{j_{l_1}}^{(1)}<u_{s_{p+1}}^{(1)} < \ldots u_{s_k}^{(1)}
\end{equation} 
where $k=N-3-l_1$.  
Similarly, if $u_{s_r}^{(2)}$ denote the values of $u$ in $\anti_f\backslash F_2$ with
$r=1,\ldots,|\anti_f\backslash F_2|$ one has the
ordering  
\begin{equation} 
u_{s_1}^{(2)}<u_{s_2}^{(2)}<\ldots u_{s_q}^{(2)}< u_{j_1}^{(2)}<u_{j_2}^{(2)} \ldots
u_{j_{l_2}}^{(2)}<u_{s_{q+1}}^{(2)} < \ldots u_{s_k}^{(2)}
\end{equation} 
where now $k=N-3-l_2$. Since the only $u$ values  below $u_1$ apart from $u=0$ must belong to
$\anti_f\backslash F_1$, $p=u_1-1$, and the   only $u$ values  below $u_2$ apart from $u=0$ and
$u_1$ must belong to $\anti_f\backslash F_1$, $q=u_2-2$ so that $p < q+1$.  

For (i) with $m>0$ and $\tm=0$, we see that $q=p+l_1-m$ so that $u_2=u_1+l_1-m+1$. For (ii) with
$m=0, \tm \geq 0$, $q=p+l_1+ \tm$ or $u_2=u_1+l_1+\tm+1$. In other words, given $u_1$, $u_2$ is
fixed to $u_2=u_1+l_1-m +\tm+1$ which covers both (i) and (ii). This relationship moreover
constrains the maximum value of $u_1$ to $N-1-l_2-l_1+m-\tm-1$. As in the case of $\anti_f=N-2$,
there is no more freedom remaining: specification of $u_1$ (along with $l_{1,2},m,\tm$ determines
the $u$ and $v$ values of the other elements uniquely.  Thus $\mu_{l_1,l_2,m,\tm}
=(N-2-l_1-l_2-m+\tm)$.  Note that the case $u_1> u_2$ is simply an automorphism (since the bulk
elements form an antichain) and hence does not contribute another factor of
$\mu_{l_1,l_2,m,\tm}$. Thus $\Psi^{(a)}_0(N-3)$ for (i) and (ii) reduce to the sums
\begin{eqnarray} 
\Psi^{(a,i)}_0(N-3) &=& A e^{-\beta R} \sum_{l_1=1}^{N-3} \sum_{l_2=1}^{N-3} \sum_{m=m_0}^{m_f}
(N-2-l_1-l_2+m) e^{\beta P m} e^{\beta Q(l_1+l_2)}\nonumber  \\  
\Psi^{(a,ii)}_0(N-3) &=& A e^{-\beta R} \sum_{l_1=1}^{(N-3-1)} \sum_{l_2=1}^{(N-3-l_1)} \sum_{\tm=0}^{(N-3-l_1-l_2)}
(N-2-l_1-l_2-\tm) e^{\beta Q(l_1+l_2)}, \nonumber 
\end{eqnarray} 
where $P=24 \epsilon^4$ and the limits for $m$ are $m_0=max(1,l_1+l_2-N+3)$ and $m_f=min(l_1,l_2)$.
While these sums are straightforward to calculate, their closed form expressions are rather lengthy.

For (b) let  $e_1 \prec e_2$ denote the elements in the chain. In this case, $F_2 \subseteq
F_1\subseteq \anti_f$. Here the abundance of the inclusive intervals is $N_1=N-1$, $N_2=l_1+1$ and
$N_3=l_2$.  Since the set of future elements of $e_1$ is $F_1\cup e_2$, and thus of cardinality $l_1+1$, $u_1
\in [1, \ldots, N-2-l_1]$, and $v_1 \in  [1, \ldots, N-2-l_1]$. An argument similar to the $\anti_f=N-2$    
case shows that $v_1=N-l_1-u_1-1$. Since the future of $e_2$ is just $F_2$ and $e_1\prec e_2$, $u_2
\in [ u_1+1, \ldots N-l_2-1]$ and $v_2=N-l_2-u_2+1$. Since $v_2>v_1$ this means that $u_2<
u_1+l_1-l_2+2$ or $u_2\in [u_1+1, \ldots, u_1+l_1-l_2+1]$. Therefore the multiplicity
$\mu_{l_1,l_2}=(l_1-l_2+1)(N-2-l_1)$ and 
\begin{equation} 
\Psi^{(b)}_0(N-3)= A e^{-\beta (R-Q)}\sum_{l_1=1}^{(N-3)}\sum_{l_2=1}^{l_1}(l_1-l_2+1)(N-2-l_1)e^{\beta Q l_1}e^{\beta T l_2}
\end{equation} 
where $T=4 \epsilon^2(1-6 \epsilon+6 \epsilon^2)$.  Again, this is straightforward to evaluate.

As is evident from this calculation, as $p$ increases, the number of distinct
configurations in the bulk increases and hence calculating the multiplicity and then
$\Psi_0(\final)$ becomes rapidly more complicated.   Instead,  following
\cite{2dqg} we now use Markov Chain Monte Carlo(MCMC) methods to generate the averaged action
$\avS_\beta(\final)$ from the equilibrium partition function and additional  numerical tools  to
obtain $\Psi_0(\anti)$.

\subsection{Numerical Calculations} 

As in \cite{2dqg} $\avS_\beta(\final)$ is obtained from MCMC simulations using a module in the
\texttt{Cactus} framework \cite{cactus}. Starting from an arbitrary 2d order $U\cap V$, the Markov
move consists of picking a distinct pair of elements from $U$ (or $V$) at random and exchanging
them.  This gives a new 2d order which we reject immediately if $\final$ is changed.  If $\final$ is
unchanged, on the other hand, the new 2d order is accepted or rejected according to the
Metropolis-Hastings algorithm for detailed balance using the relative weights $\exp(-\beta
S_{\beta}(\final))$ of the initial and final causal sets.  Using various observables including the
action itself, this process is seen to thermalise fairly rapidly for most values of $\final$. From
the thermalised ensemble, the average of the action $\avS_\beta(\final)$ is obtained. In order to
calculate 
\begin{equation} 
\Psi_0(\final) = A \partn_\beta(\final) = A \partn_0(\final)
\exp({-\int_0^\beta d\beta'  \avS_{\beta'}(\final)})
\end{equation} 
it is therefore also necessary to obtain the partition function $\partn_0(\final)$  at $\beta=0$, 
and  perform the above  numerical integration of $\avS_{\beta'}(\final)$ over $\beta'$.
 
We restrict our discussions to simulations for $N=50$ element 2d orders, with
$\epsilon=0.12,0.5,1.0$. Less extensive simulations for different $N$ suggest
that these results are robust. Indeed, the interplay between $N,\beta, \epsilon$ gives rise to a
non-trivial scaling behaviour when $\infin$  is unrestricted,  indicative of a well defined asymptotic
limit \cite{2drg}. We expect the same to be true for fixed $\infin$ but a discussion of this is
beyond the scope of the current work.

\subsubsection{Calculating $\mc{Z}_0$}

The partition function $\partn_0$ with no restrictions on $\infin$ can be written as $\partn_0=
\!\sum_{\final} \partn_0(\final)$, where $\partn_0(\final)$ is the restricted partition function.
Since $\partn_0$ is dominated by 2d random orders $\partn_0(\final)$ is given (up to overall
normalisation) by the frequency of those with fixed $\final$ from an ensemble of 2d random orders.
We simulate $1.138\times 10^{10}$ 2d random orders to generate this frequency profile for $\final$
up to $19$ and find
\[
\begin{array}{rr}
N_f& \text{No of occurrences}\\
 1 & 2.32246\times 10^8 \\
 2 & 1.03553\times 10^9 \\
 3 & 2.12\times 10^9 \\
 4 & 2.68325\times 10^9 \\
 5 & 2.37839\times 10^9 \\
 6 & 1.58286\times 10^9 \\
 7 & 8.27049\times 10^8 \\
 8 & 3.50007\times 10^8 \\
 9 & 1.22738\times 10^8 \\
 10 & 3.62907\times 10^7 \\
 11 & 9.1807\times 10^6 \\
 12 & 2.00639\times 10^6 \\
 13 & 383502. \\
 14 & 63963. \\
 15 & 9483. \\
 16 & 1191. \\
 17 & 146. \\
 18 & 22. \\
 19 & 3. \\
\end{array}
\]
Despite the large number of trials we only found an $N_f=19$ causet three times and no causet with larger $N_f$.
This shows how much entropy disfavors these states. Beyond this the frequency becomes numerically
insignificant. 

\begin{figure}
\centering{\includegraphics[width=0.5\textwidth]{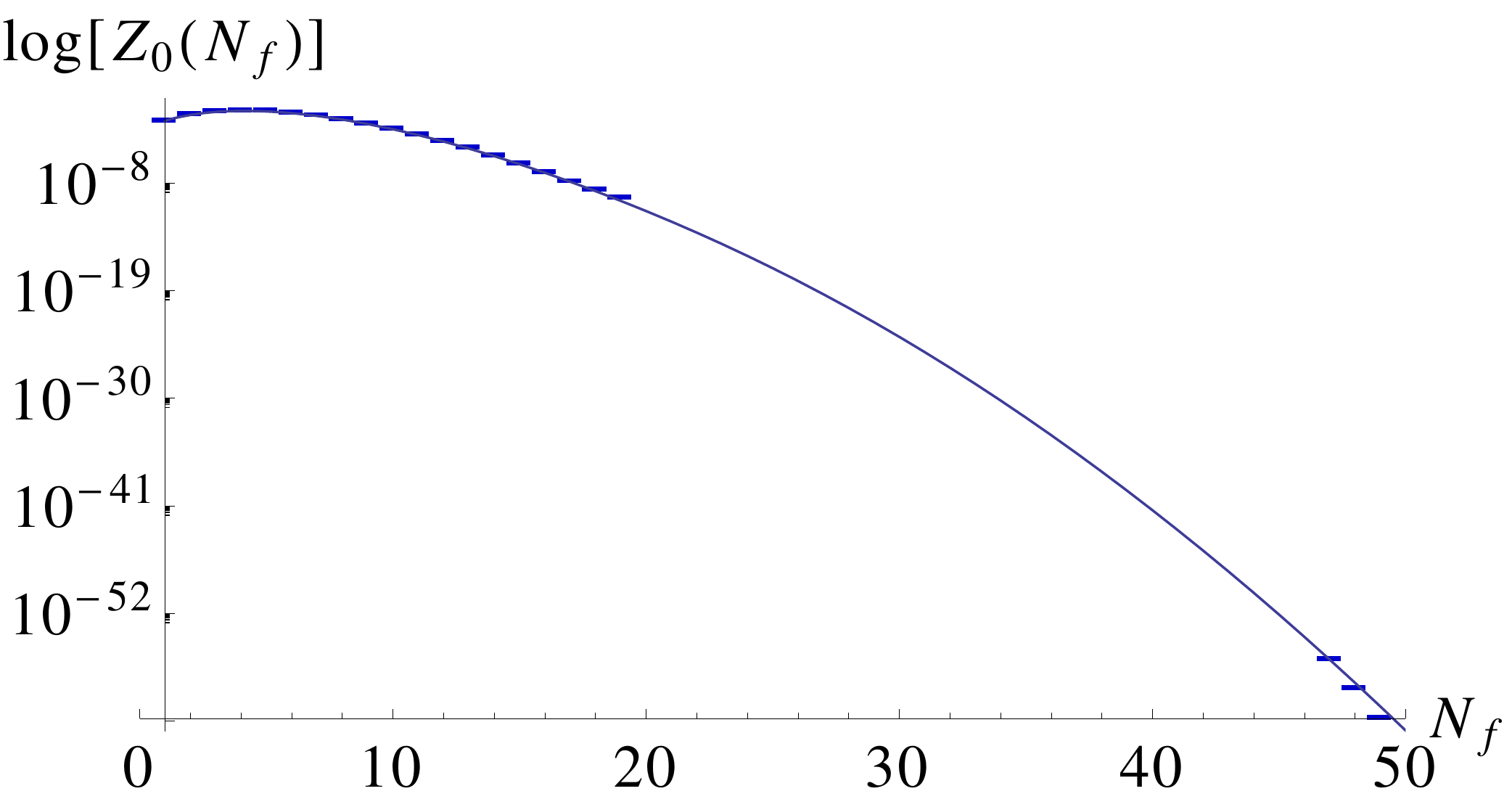}}
\caption{A log plot of the  histogram for $\final$ using $1.138\times 10^{10}$ random 2d orders.}
\label{fig:partnzero}
\end{figure}

\begin{figure} 
\subfigure[\label{fig:zzerotwo} Peak region]{\includegraphics[width=0.32\textwidth]{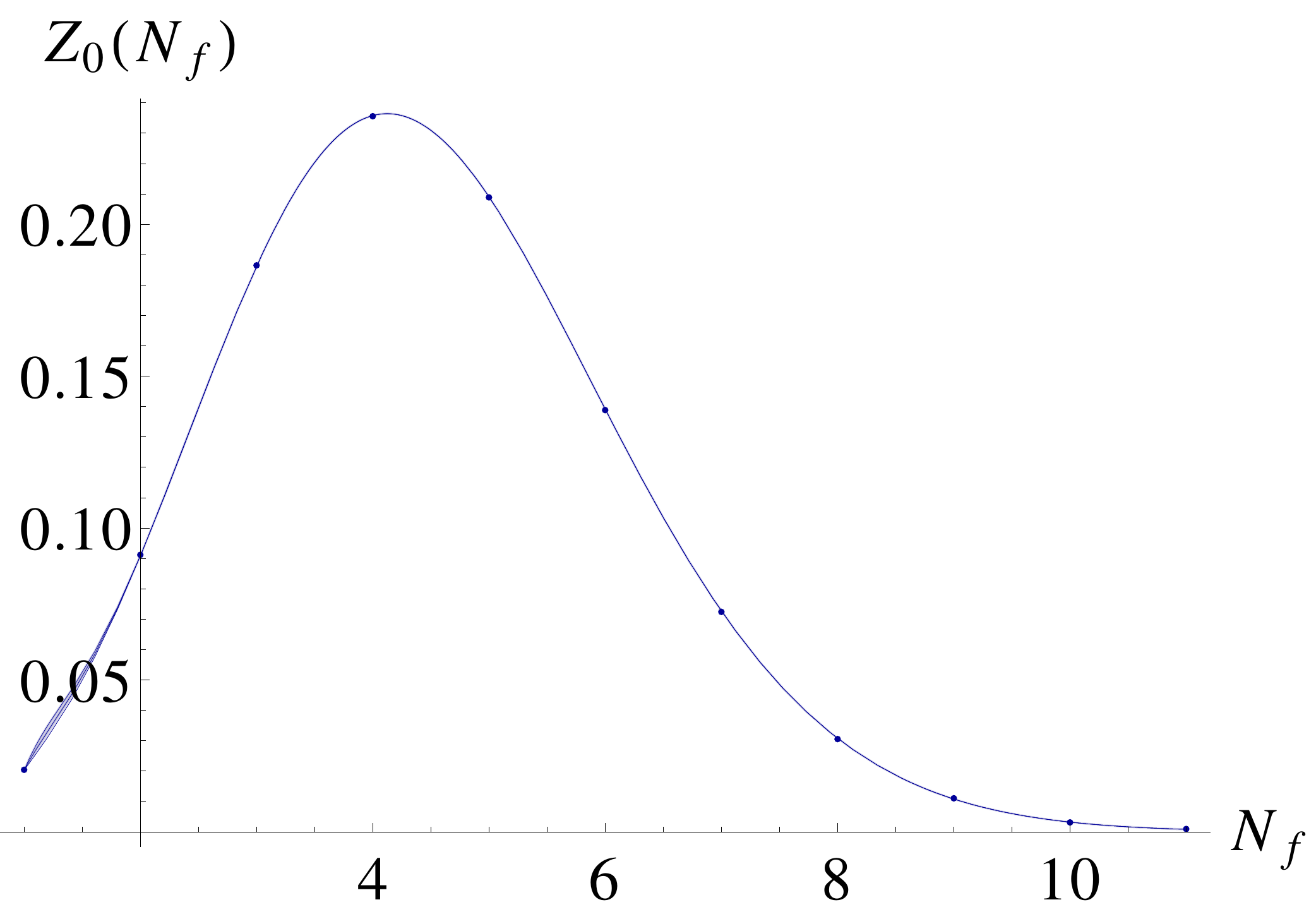}}
\subfigure[\label{fig:zzero} Error bands]{\includegraphics[width=0.32\textwidth]{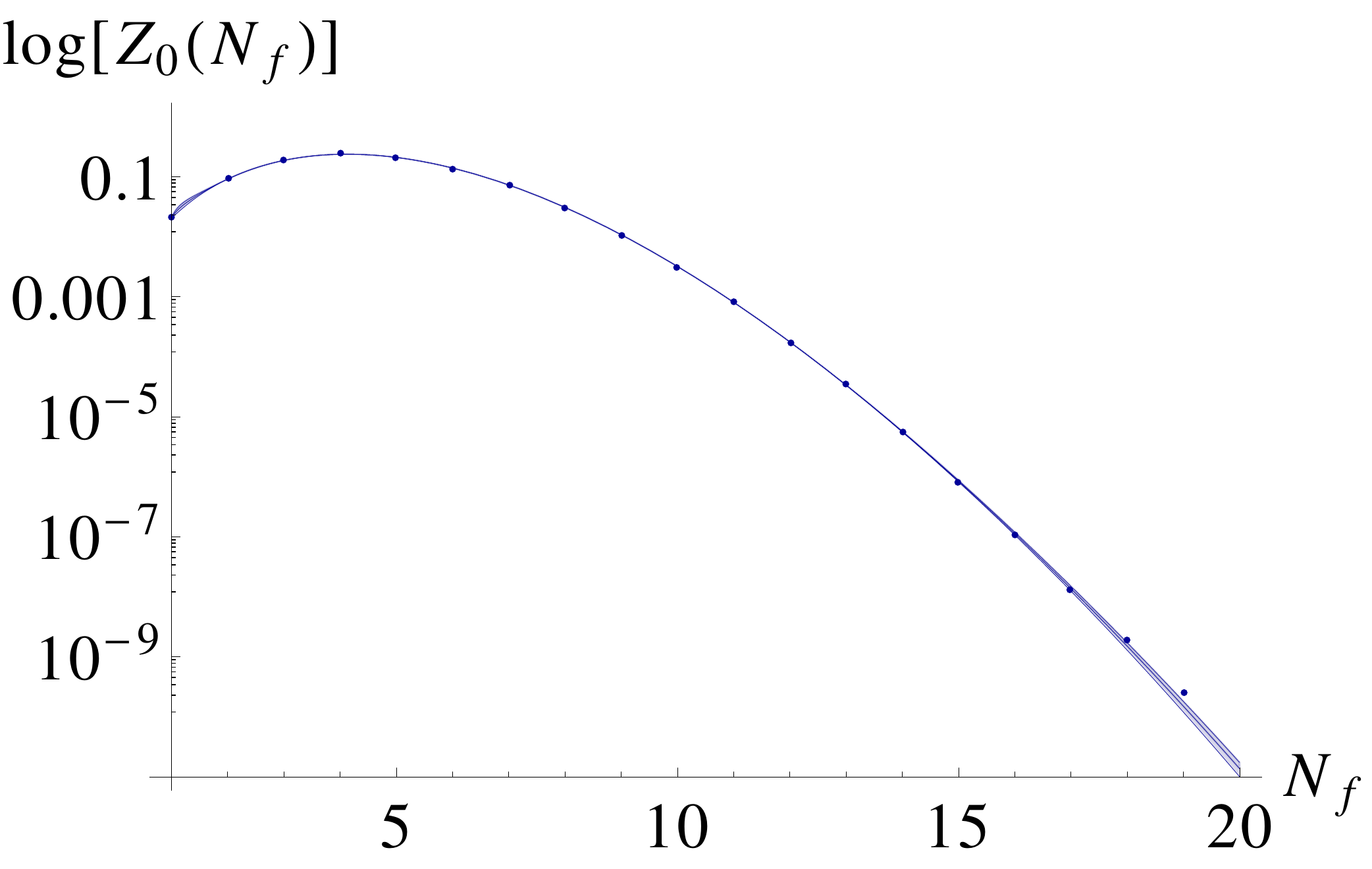}}
\subfigure[\label{fig:zzerothree}Difference to bounds and residue]{\includegraphics[width=0.32\textwidth]{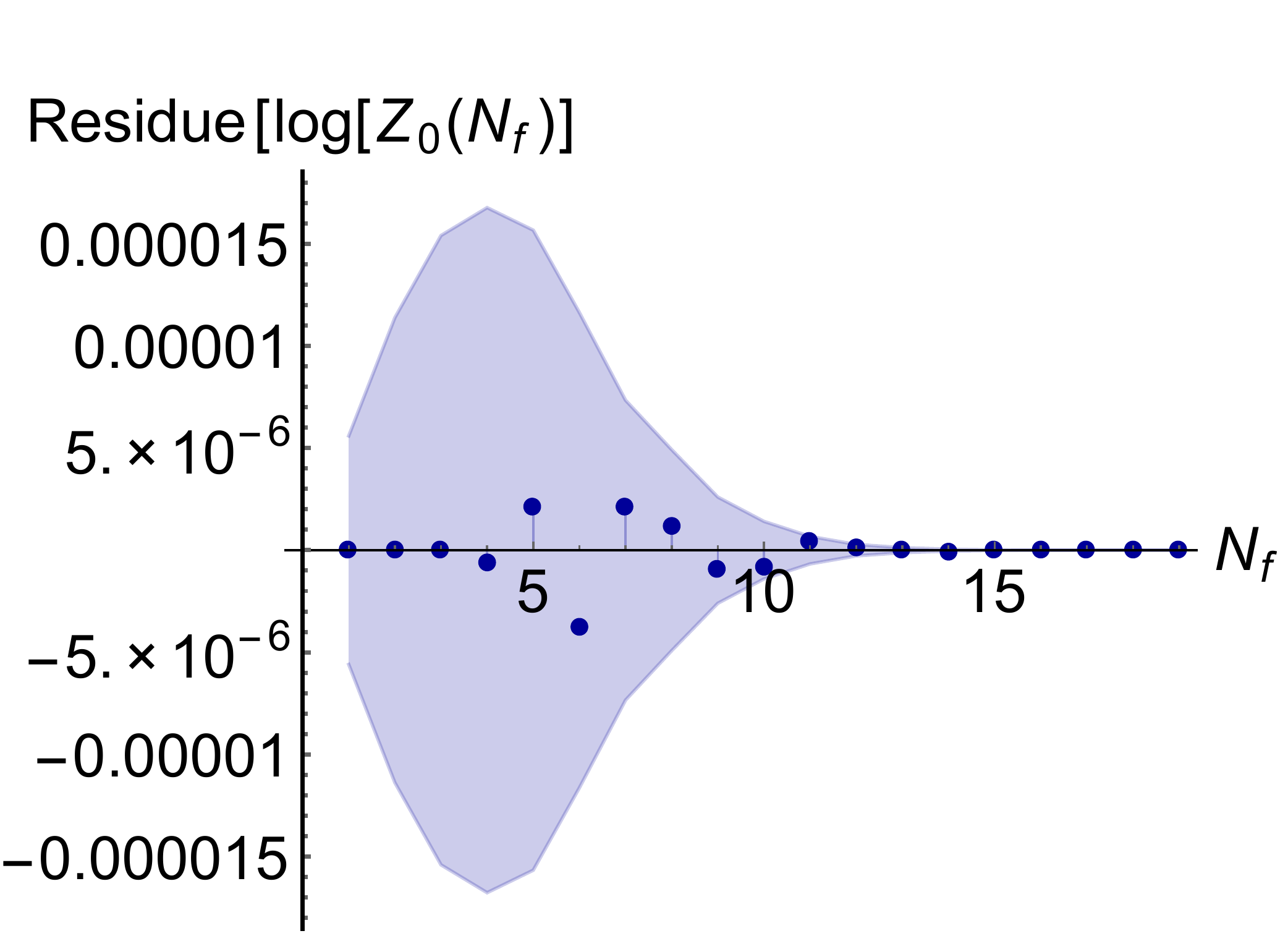}}
\caption{These plots illustrate the fitting for $\partn_0(\final)$. In Figure a) Zoom of best fit to
  $\partn_0(\final)$ with $90 \%$ confidence bands. Only the peak region is shown, on a non-log
  scale. Figure b) is the best fit to $\partn_0(\final)$ with $90 \%$ confidence bands. The errors are very
  small and  only visible in the lower right corner. In Figure c) the dots are the difference between the calculated
value for Z 0 (N f ) and the measured value, while the blue shaded region shows the upper and lower
bounds for the estimate.
}  
\end{figure}

Next, we use these data points together with the analytic results for $N_f=47,48,49$ to fit a
function for $\partn_0(\final)$. The  best fit estimate is shown in Figure \ref{fig:partnzero}.
The fit function is determined by best guess as
\begin{align}
(a + h x^m + (e + f x + g x^2 + j x^3 + k x^{5.5}) \ln(x)) e^{-b (x + d)^2)}
\end{align}
The large number of free parameters is not important, since we only need a function that generates a
good estimate of the values.  The fit works very well and the error bands on it, as determined by
\texttt{Mathematica}, are so small that we have difficulty in showing them in a plot.  The plot
Figure \ref{fig:zzero} shows that the errors are visible in the lower right corner only. Since the
small error might seem an effect of the log plot and the choice of region we zoom into the peak of
the distribution in a non-log plot Figure \ref{fig:zzerotwo}.  To see the very small errors, we
include Figure \ref{fig:zzerothree}, which shows the difference between the estimate for
$\partn_0(\final)$ and the real values together with the uncertainty on the estimate.  The quality
of the estimate is therefore very good and we use it as the function $\partn_0(\final)$ for the
remaining part of the analysis.

\subsubsection{MCMC simulations} 

\begin{figure}
{\centering 
\subfigure[Thermalisation for $\final=15$, $\beta=1$, $\epsilon=0.12$.]{
\includegraphics[width=0.7\textwidth]{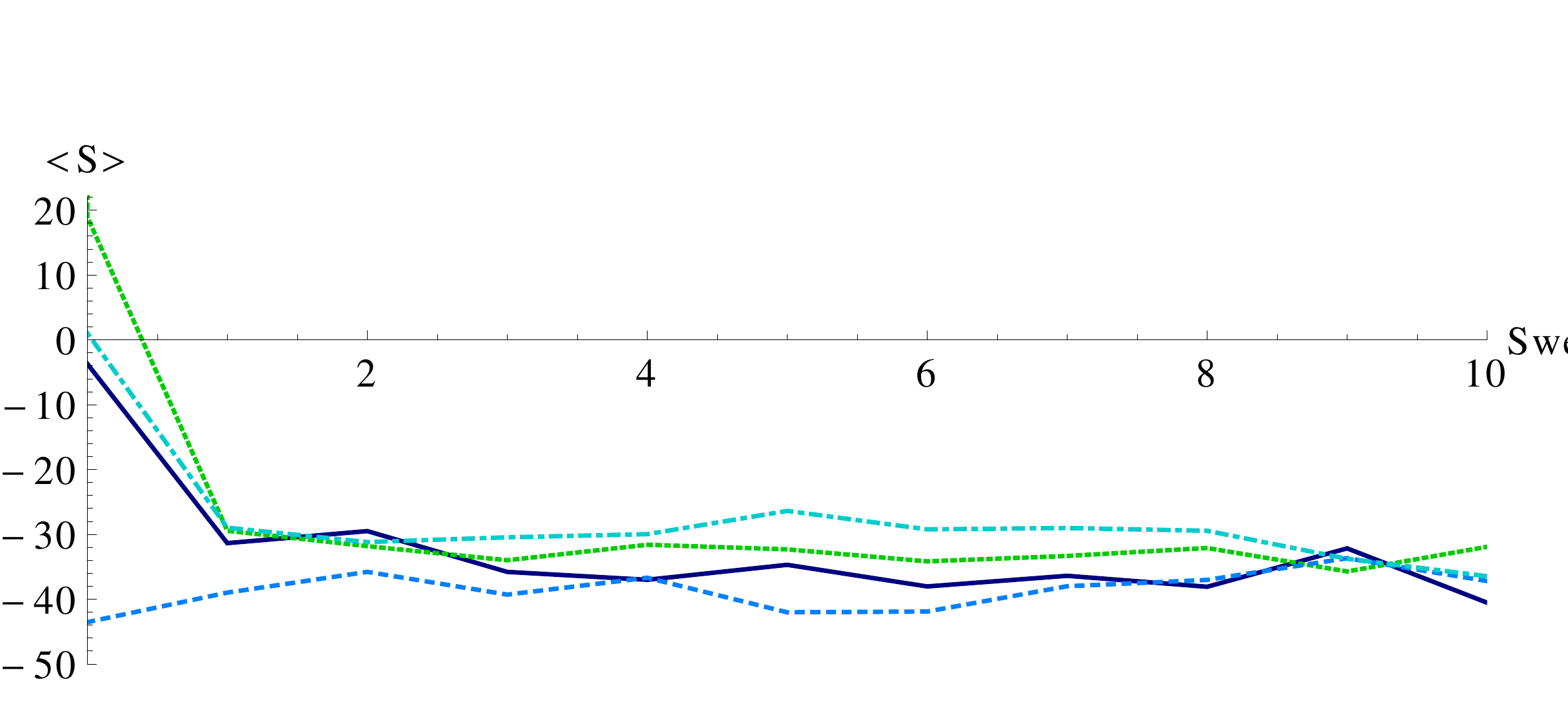}\hspace{10pt}}
}
\subfigure[Average action starting from different configurations for $\beta=7.6$ $N_f=27$]{\includegraphics[width=0.5\textwidth]{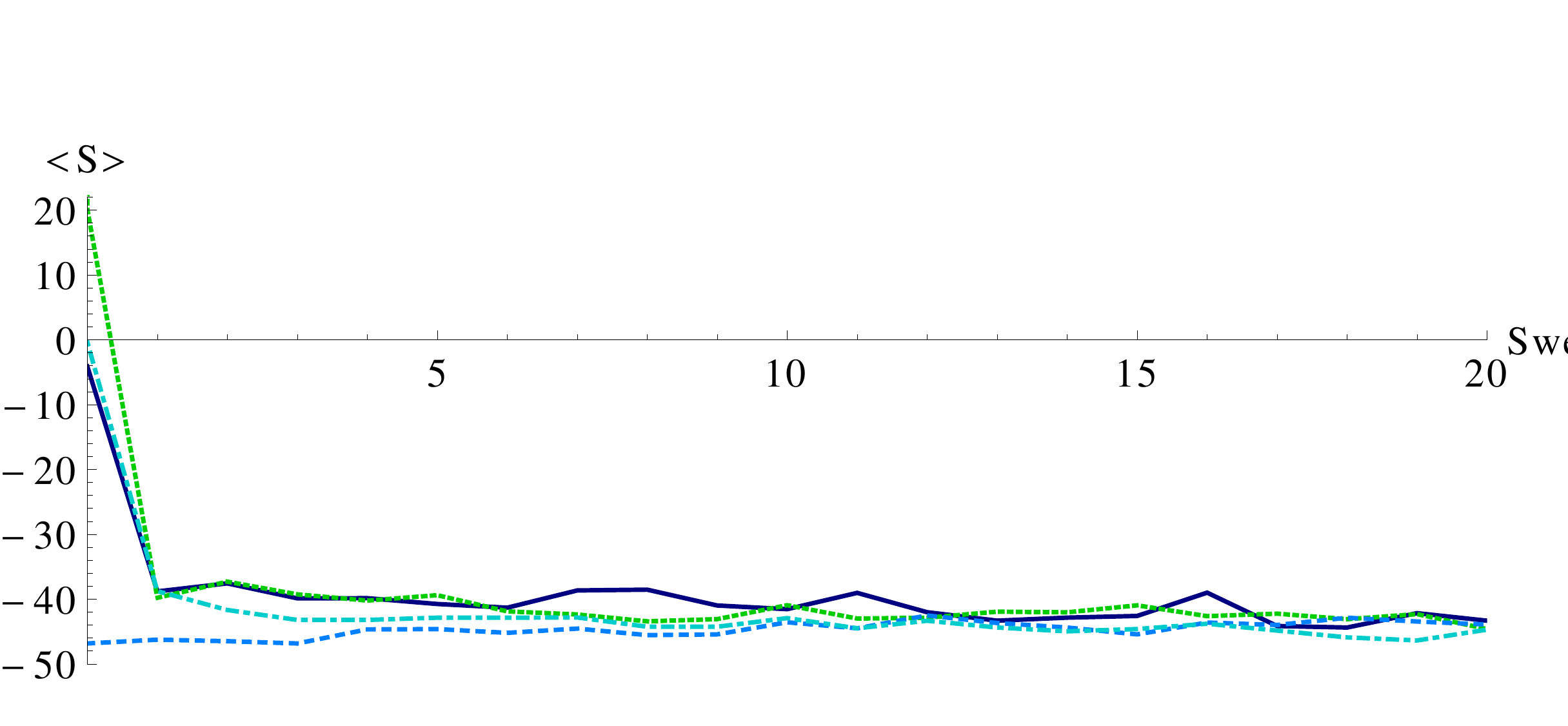}}
\subfigure[Average action starting from different configurations for $\beta=6.8$ $N_f=45$]{\includegraphics[width=0.5\textwidth]{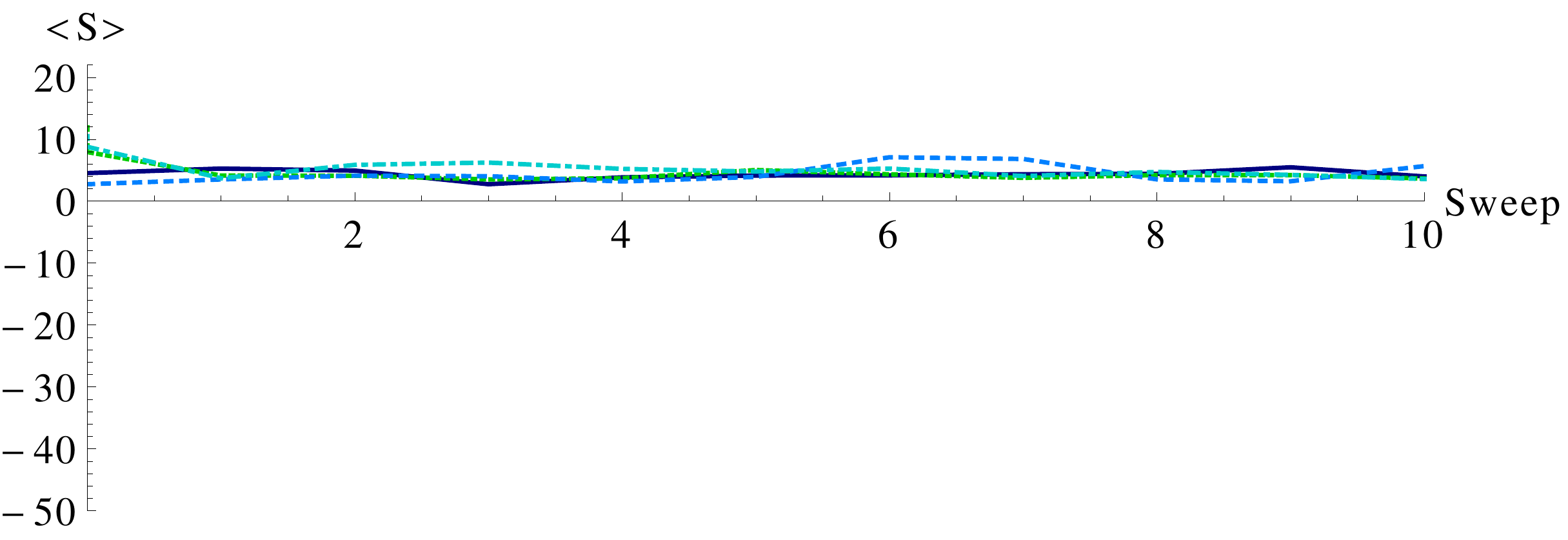}}
\subfigure[Average action starting from different configurations $\beta=6$ $N_f=30$]{\includegraphics[width=0.5\textwidth]{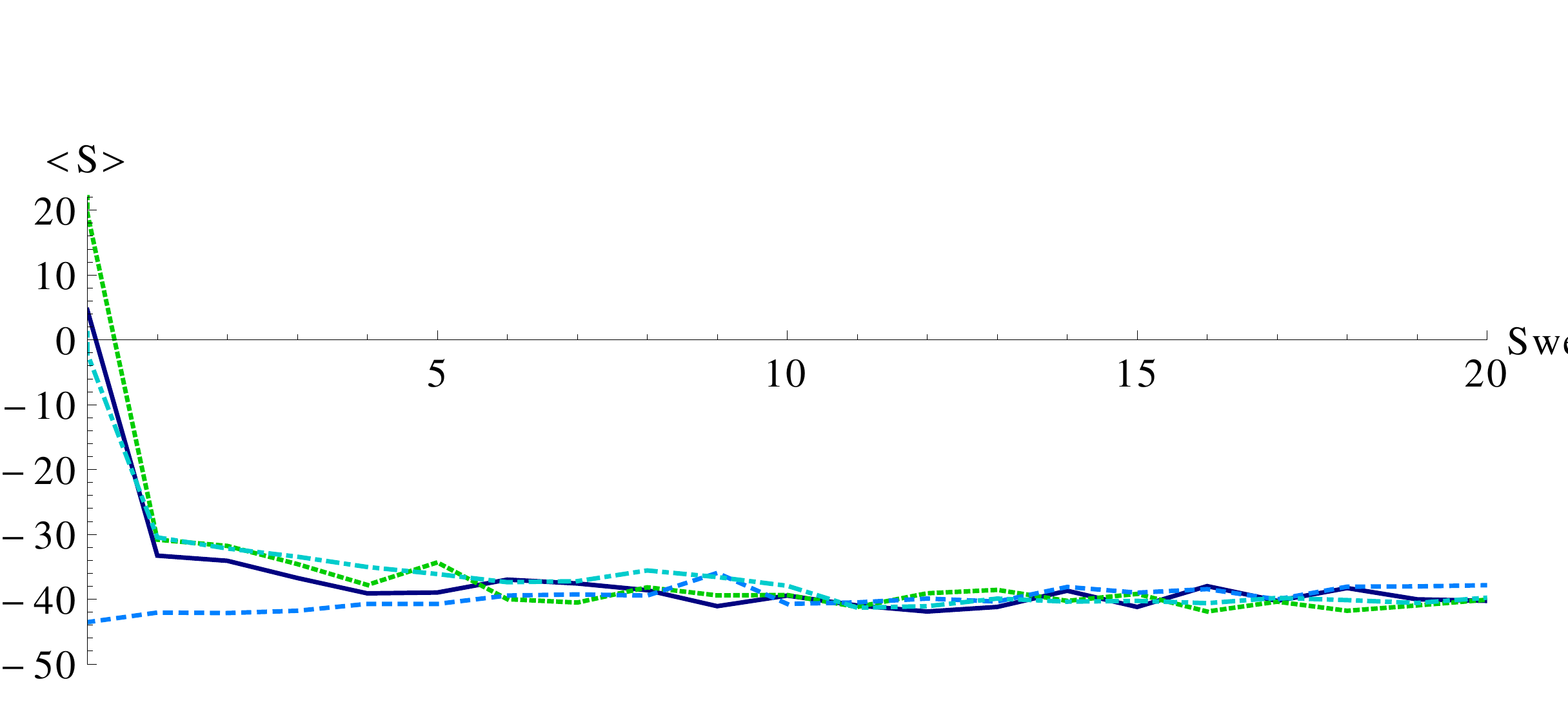}}
\subfigure[Average action starting from different configurations $\beta=4$ $N_f=30$]{\includegraphics[width=0.5\textwidth]{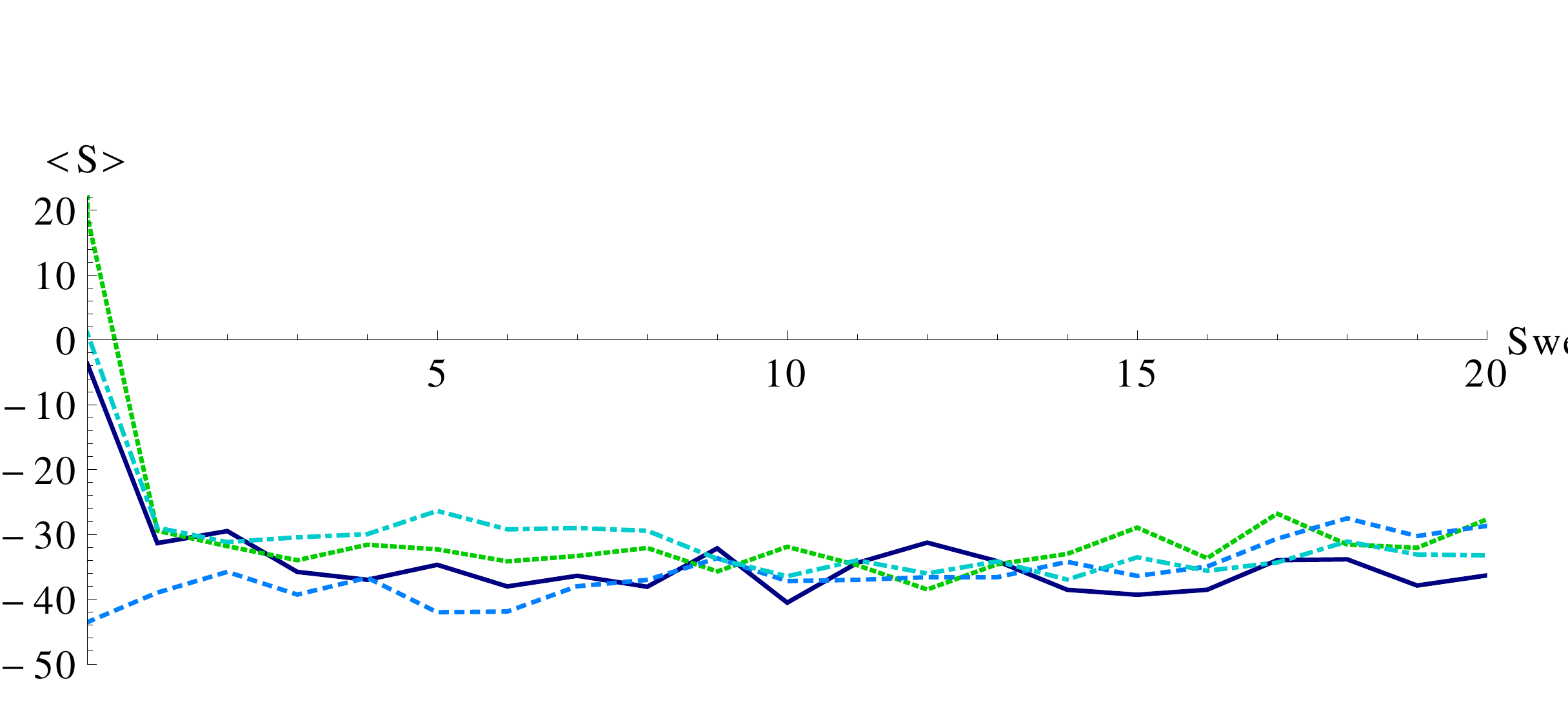}}
\caption{\label{fig:therm}Starting from different initial configurations thermalisation is reached quickly.}
\end{figure}

For the MCMC simulations we define a sweep as $\binom{N}{2}$ moves, and perform $10,000$ sweeps for
$\final=1,\dots 46$. To compare with the best data in \cite{2dqg} the $\epsilon=0.12$ trials are
done in finer steps of $\beta$ and coarser steps for $\epsilon=0.5,1$. In all three cases, the
qualitative features observed are the same.  Since thermalisation problems set in at different
values of $\beta$ for different $\epsilon$, for $\epsilon=0.5,1$, the data, though coarser, spans
more of the $\beta>\beta_c$ regime.  Our conclusions take all this data into account.
Analytic results are used for $\final=47,48,49$. Thermalisation typically occurs very quickly,  an example of which is
given in Fig (\ref{fig:therm}).
We give a more detailed picture of the thermalisation below 

\noindent {{\bf Tests of thermalisation}}\\
To ensure that our simulations thermalise we start from different bulk configurations that lie
between the  initial and final antichain.
\begin{itemize}
\item a total chain
\item a total antichain
\item a random 2d order
\item a crystalline order
\end{itemize}

Our code also allows us to start the simulations from a given configuration in a file. 
This makes it possible to resume simulations in a thermalised state, or test the thermalisation of
special configurations. 
To test the thermalisation of the configurations used in our analysis we ran the code starting from different initial
conditions for the three values of $\epsilon$ varying over $\beta$. 
We deemed the thermalisation to be sufficient if the average action for the different initial
configurations agreed to within the error bars.

We found that thermalisation properties are fairly good.
For $50$ element causets the configuration thermalises after few moves, independent of the length of
the final chain, 
for smaller values of $\beta$, but gets slower as  $\beta$ increases.  
In Figure \ref{fig:therm} we show the thermalisation for some examples of large $\beta$ and
moderately to large $N_f$.  These configurations are those that might have had  the most problems
with thermalisation. Yet,  as shown here, they in fact thermalise very fast.

\begin{figure}
\centering{\includegraphics[width=0.7\textwidth]{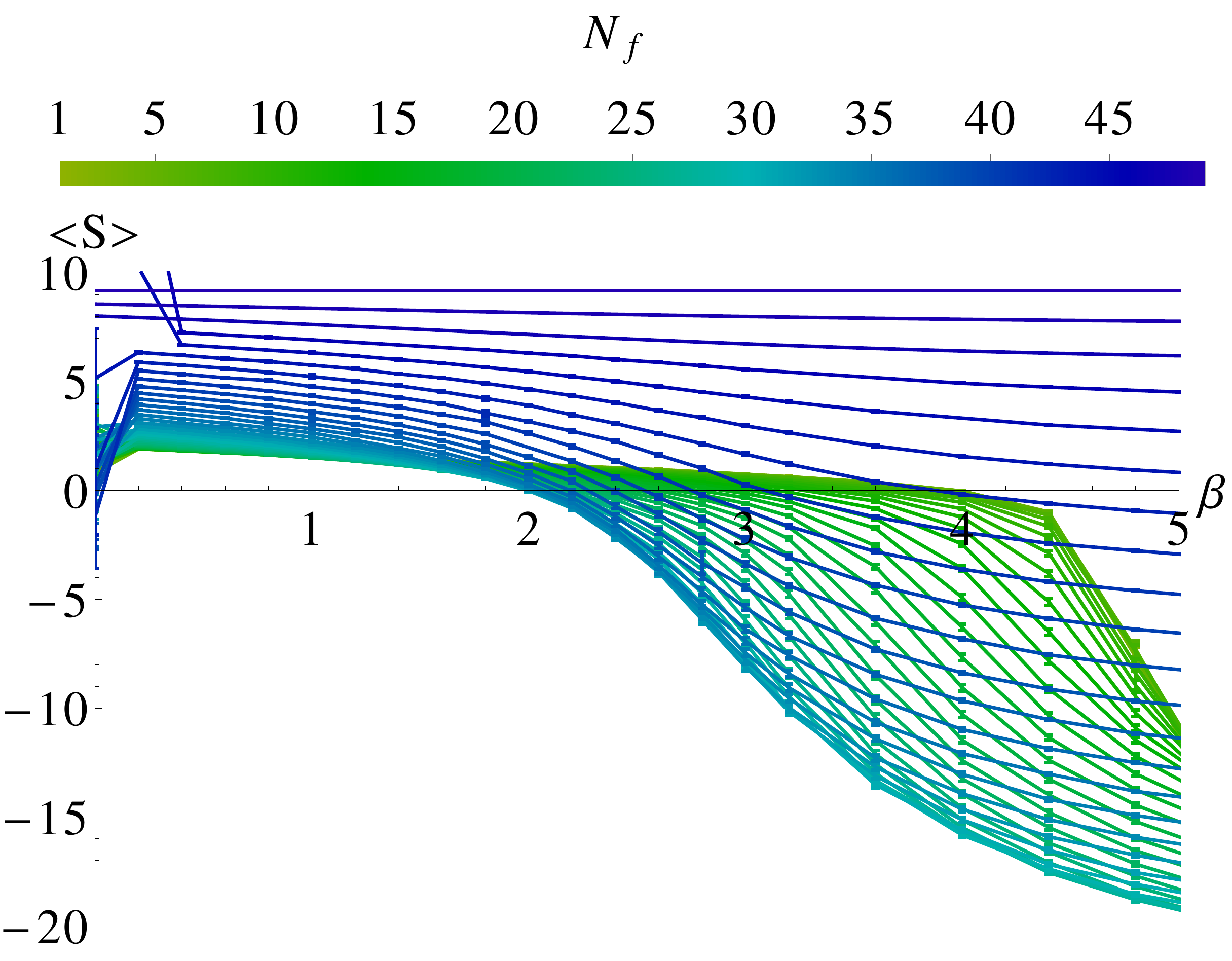}\hspace{10pt}}
\caption{$\avS_\beta(\final)$ as a function of $\beta$,  $\epsilon=0.12$.  }
\label{fig:expS}
\end{figure}

\noindent {{\bf Results of MCMC simulations}} \\ 
The $\final$ dependence of the phase transition is shown in Figure \ref{fig:expS}.
For small values of $\final$ the behaviour of the phase transition is similar to that in
\cite{2dqg}  with a continuum phase for $\beta<\beta_c$ and a crystalline phase
for $\beta>\beta_c$. As $\final$ increases the critical point $\beta_c$ first begins to decrease
achieving a minimum value $\beta_c^{min}$  around $\final \sim 30$ after which it begins to increase again. For
$\final > 40$, the nature of the transition changes since a reduced bulk makes the two phases less
distinguishable. 

This   rich phase structure that emerges from our calculations contains the essence of what the
following analysis  will extract. In particular, one can with the eye begin to see the reason for the
dominance of certain configurations over others, based on the temperature at which the particular phase
transition sets in.

\subsubsection{Numerical Integration} 

\begin{figure}
\centering{\includegraphics[width=0.7\textwidth]{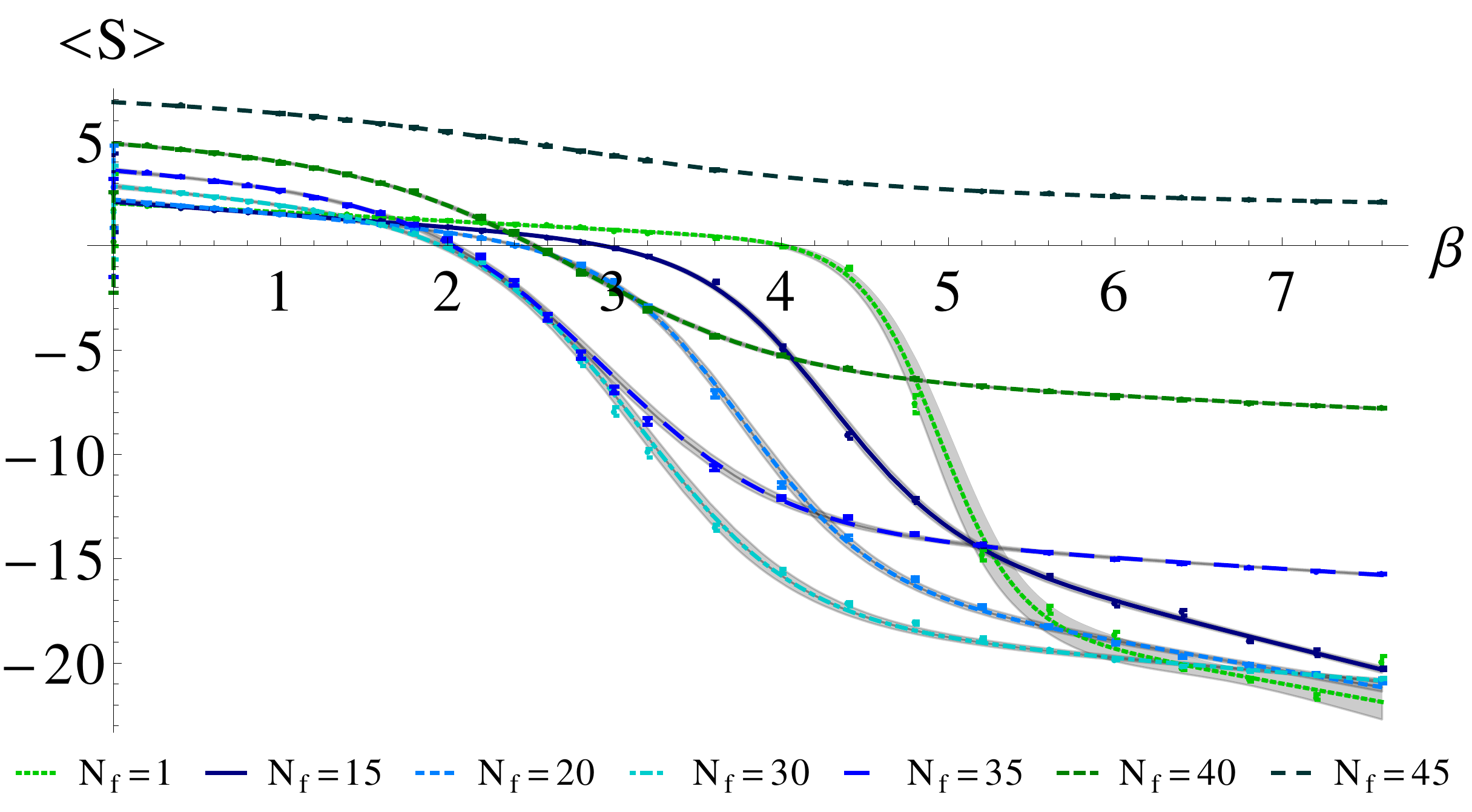}\hspace{10pt}}
\caption{Errors in the interpolating functions for $\avS_\beta(\final)$ as a function of $\beta$, $\epsilon=0.12$. }
\label{fig:actionerrors}
\end{figure}

\begin{figure}
\subfigure[$\av{\Sa(\beta,0.12)}$ with best fit]{\includegraphics[width=0.5\textwidth]{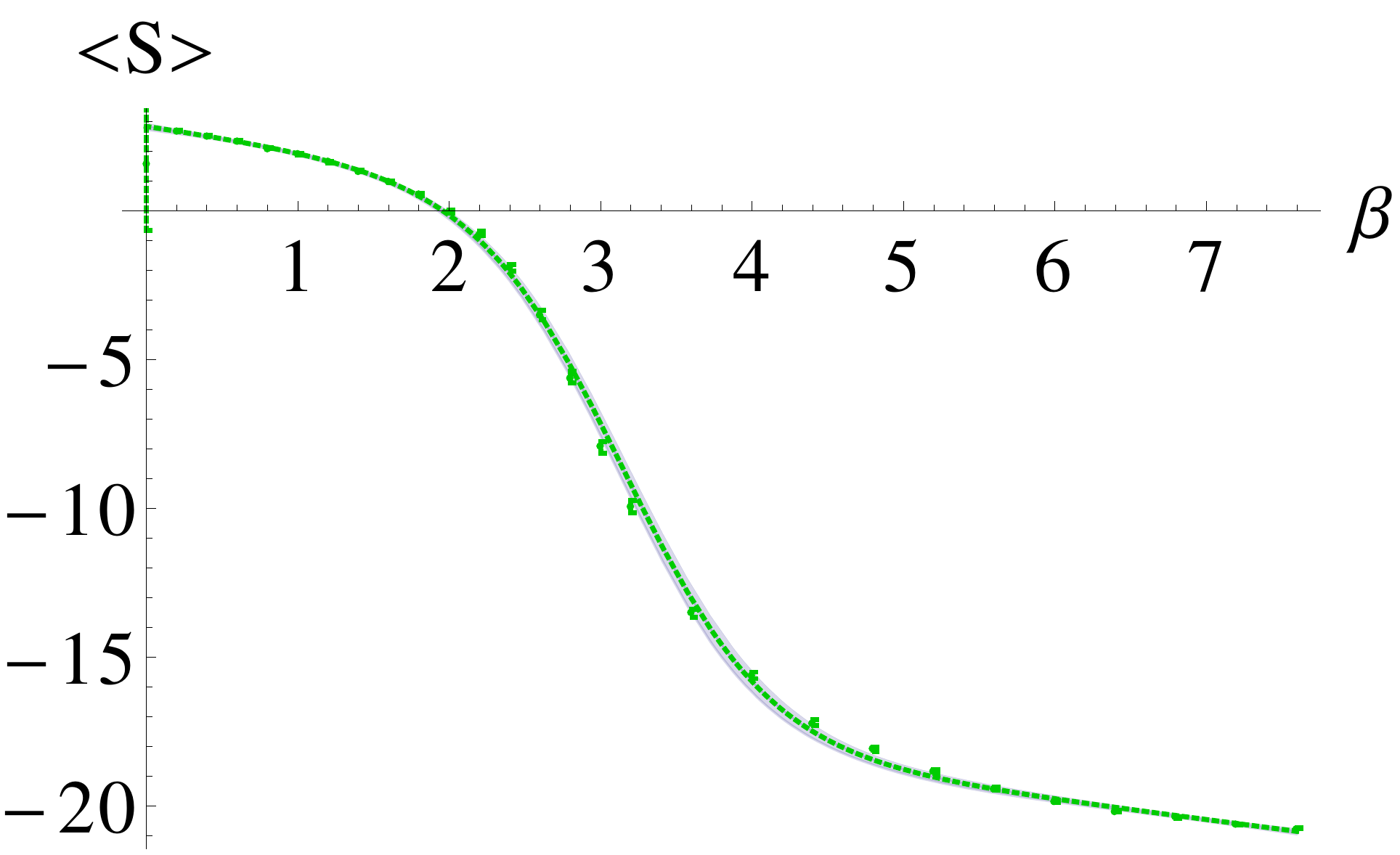}}
\subfigure[$-\ln{\partn_{\beta}(30,50)}$  ]{\includegraphics[width=0.5\textwidth]{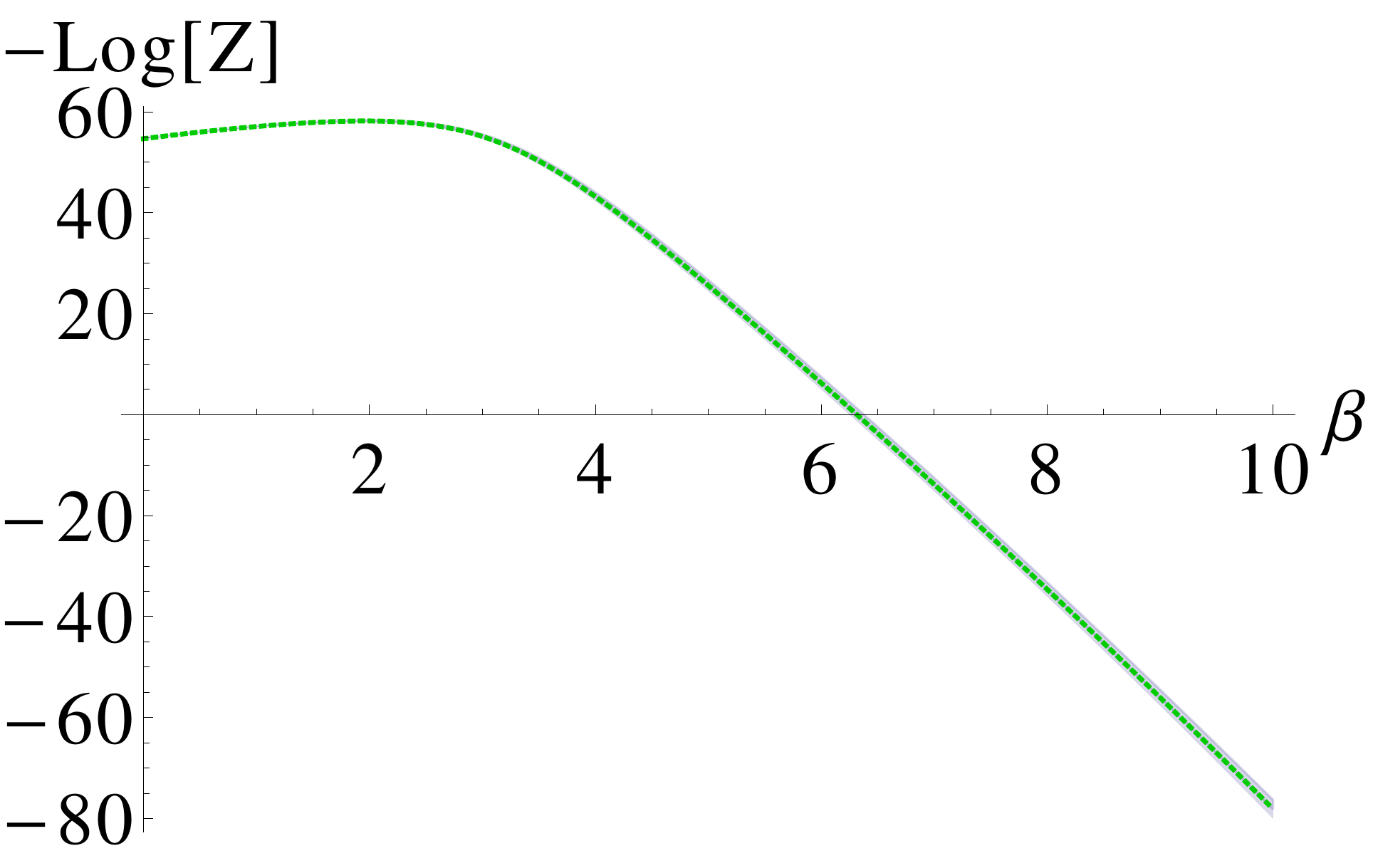}}
\caption{\label{fig:casestudy} 
The figure on the right, $-\ln{\partn_{\beta}(30)}$ , is calculated by integrating the left hand figure,$\av{\Sa(\beta,0.12)}$, and subtracting $-\ln{\partn_{0}(30)}$.}
\end{figure}

Our MC simulations give the average action $\avS_\beta(\final)$. We want to numerically integrate it  to find
\begin{align}
\ln {\frac{\partn_0(\final)}{\partn_\beta(\final)}}=
\int_0^\beta d\beta'  \avS_{\beta'}(\final)
\end{align}
To evaluate the RHS of this equation, we begin with tabulating our measurements of $\beta$ and $
\av{\Sa(\beta,\epsilon)}$.  We then make a best fit for this data and numerically integrate it
using \texttt{Mathematica}.  For $\epsilon=0.12$ we use the function 
\begin{align}\label{eq:fitfun}
(a + f\,  x) \tanh(b x + c) + d + e x
\end{align}
Using a best fit function instead of an interpolation between the points does allow us to use the
additional data contained in the measurement errors of the average action.  The best fit also leads
to a smoother and more consistent estimate of $\partn_\beta(\final)$ compared to using a pure
interpolation function. To demonstrate our method of calculation we show show the $\final=30$ case
in detail. On the left hand side of Figure \ref{fig:casestudy} we show $\avS_\beta(\final=30)$ for
$\epsilon=0.12$ together with the error bars, the fitted function and the shaded region.  On the
right hand side we show $-\ln{\partn_\beta (30)}$ calculated by integrating average action and
subtracting $-\ln{\partn_0(30)}$.  If we had approximated the average action through line segments
the result for $-\ln{\partn_\beta(30)}$ would show jumps at the points where two line segments meet,
especially at the beginning and the end of the phase transition.
Importantly, introducing  several free parameters in fitting the function is not physically
significant, since the fit parameters are not themselves of independent interest.

For $\epsilon=0.5$ we need to use a different fit function 
\begin{align}
(a+b x)\Theta(-x+c)+d \Theta(x-c)
\label{eq:stepfit} 
\end{align}
where $\Theta$ is the Heaviside step function.

\subsubsection{The HH Wavefunction} 
Putting together the estimate of $\partn_0(\final)$ with the above results of numerical integration,
we can finally normalise $\Psi_0(\final)$ using $\Sigma_{\final=1}^{N-1}
|\Psi_0(\final)|^2=1$. This gives us $\Psi_0(\final)$ as a function of $\beta$. What is very
surprising is that rather than obeying a fairly generic behaviour, $\Psi_0(\final)$ displays clear
peaks about specific discrete geometries.  A careful examination shows that this is a result of the
rich phase structure displayed in  Fig \ref{fig:expS} and the existence of a value of $\final$ at
which the critical $\beta$ is the smallest.   

\begin{figure}
\subfigure[$\epsilon=0.12$]{\includegraphics[width=0.5\textwidth]{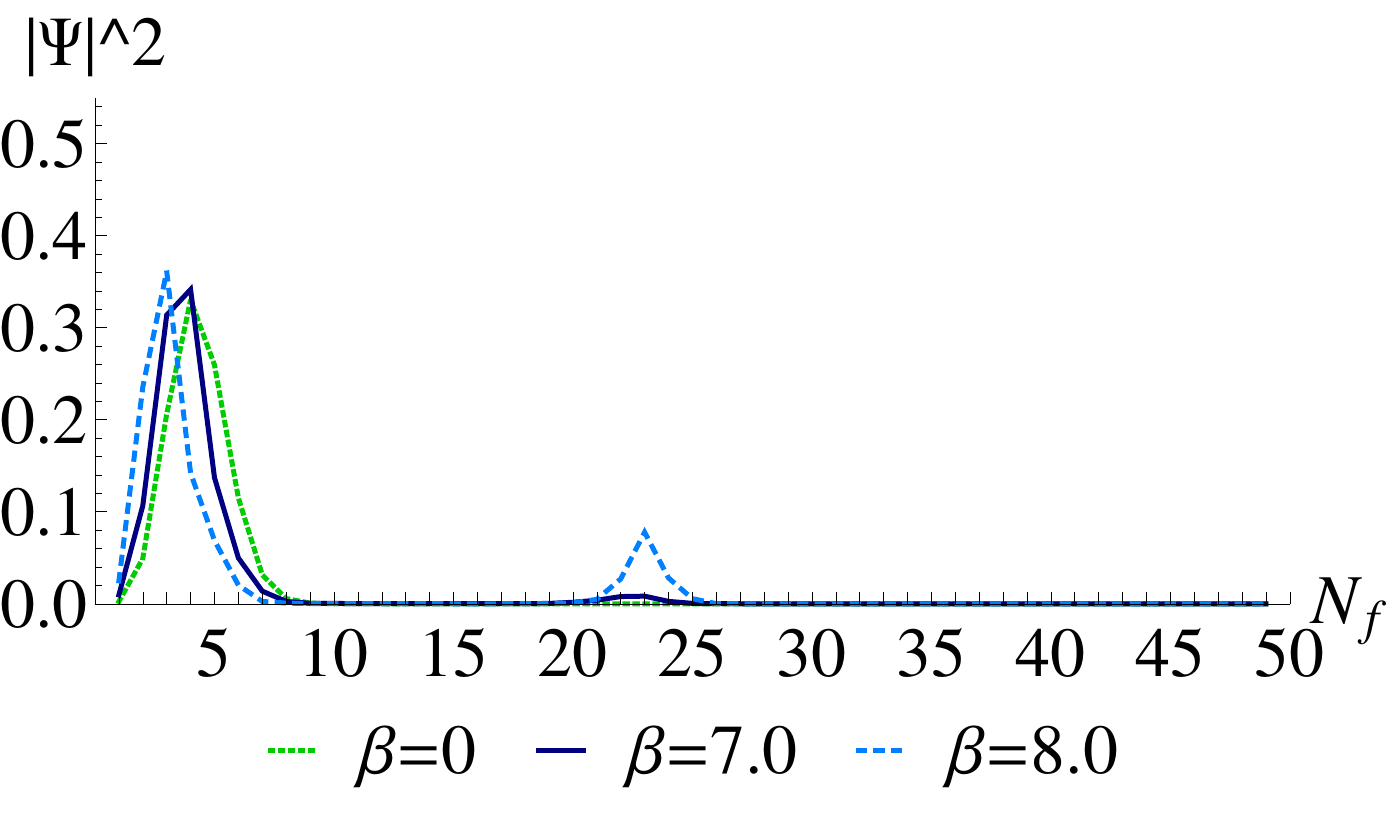}} 
\subfigure[$\epsilon=0.5
$]{ \includegraphics[width=0.5\textwidth]{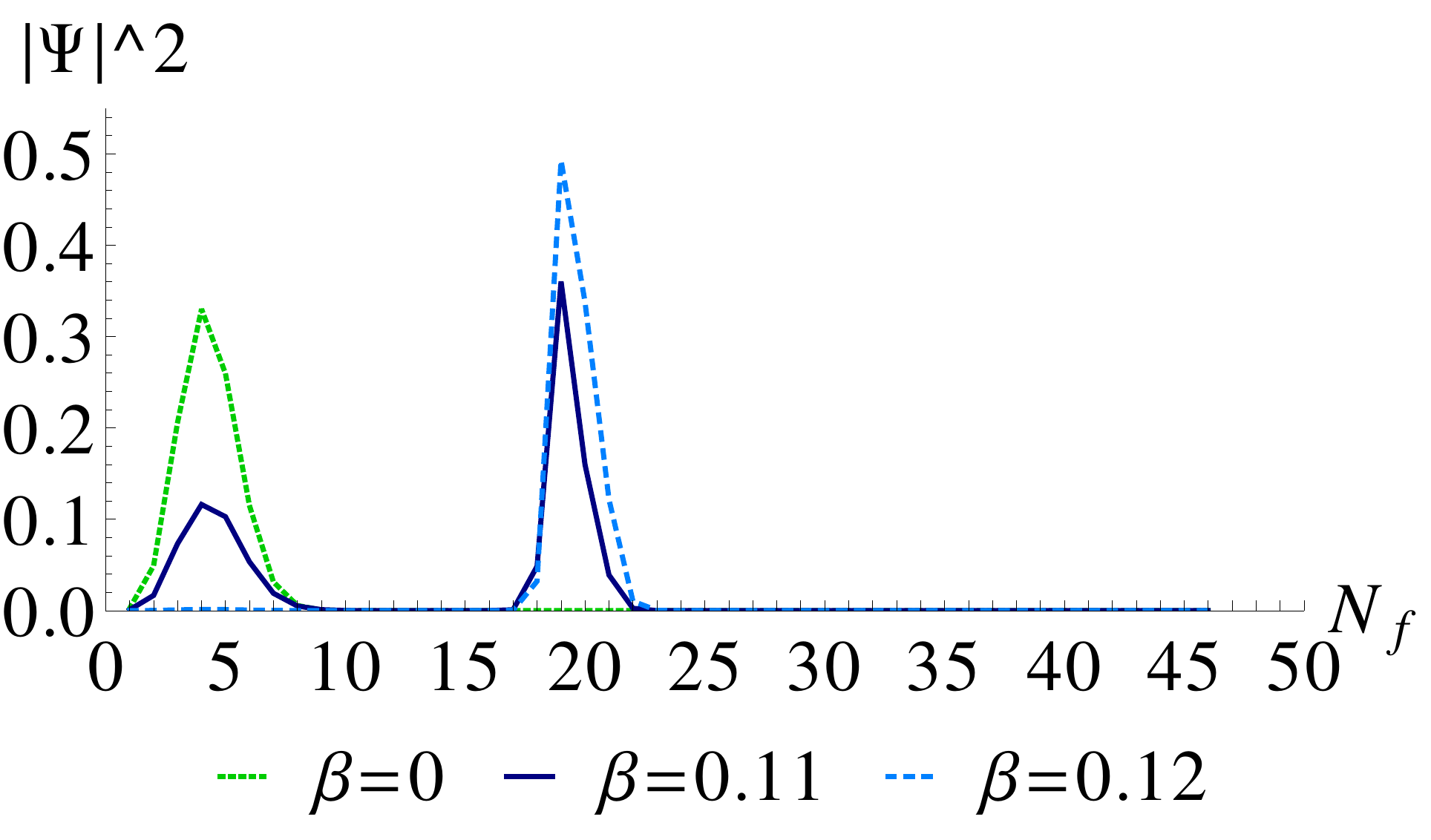}}
\caption{
  $|\Psi_0(\final)|^2$  for
  $\epsilon=0.12,0.5$}
\label{fig:psialone}
\end{figure}

As $\beta$ increases, moreover, there is the interesting struggle displayed in $\Psi_0$ between the
``entropic'' component, $\partn_0(\final)$, and the action. This is shown in 
Figure \ref{fig:psialone} for $\epsilon=0.12,0.5$.
For small $\beta$ it is dominated by the entropic contribution and is peaked around $\final \sim 4$
starting at $\beta=0$ all the way upto $\beta \sim 7.6$.  Around $\beta \sim 7.0$
$|\Psi_0(\final)|^2$ develops a second peak at $\final \sim 23$ which gets more pronounced as
$\beta$ increases.  Though the thermalisation properties of the data begin to deteriorate beyond
$\beta \sim 8$ there is an indication that the second peak continues to grow and the first peak
shrinks.  This shifting of peaks also occurs for $\epsilon=0.5$ and $\epsilon=1$; for these the
second peak clearly begins to dominate the first as one goes to larger $\beta$ as shown in Figure
\ref{fig:psialone}.  Hence it appears that as $\beta$ goes well past $\beta_c^{min}$ the second peak
dominates the first. Importantly, the existence of well formed peaks at all $\beta$ does not arise
from tweaking of parameters, but from the details of the phase transitions seen in Figure
\ref{fig:expS}.

\begin{figure}
\centering
\subfigure[$\beta=0$]{\includegraphics[width=0.3\textwidth]{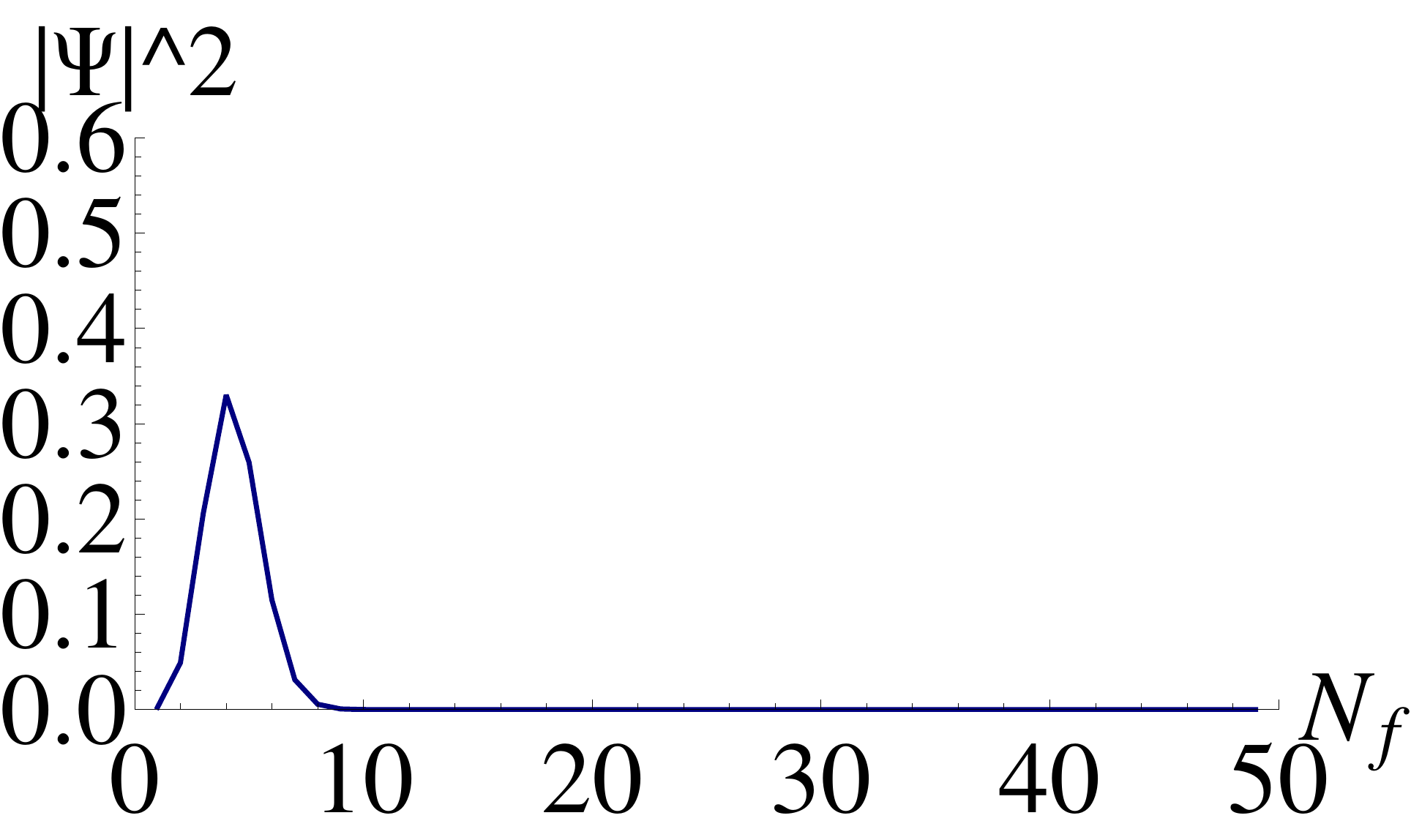}}
\subfigure[$\beta=7.5,\epsilon=0.12$]{\includegraphics[width=0.3\textwidth]{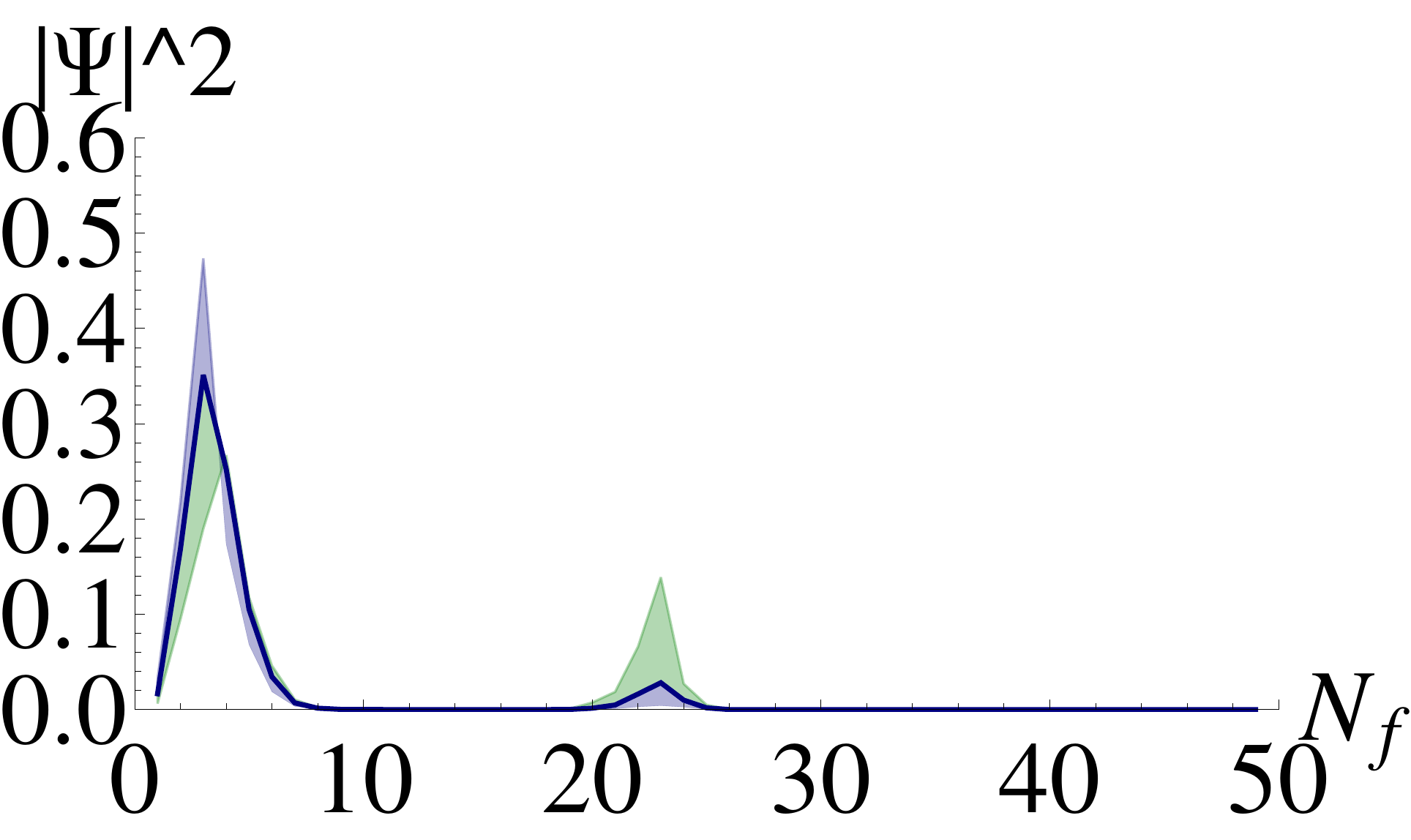}}
\subfigure[$\beta=8.5,\epsilon=0.12$]{\includegraphics[width=0.3\textwidth]{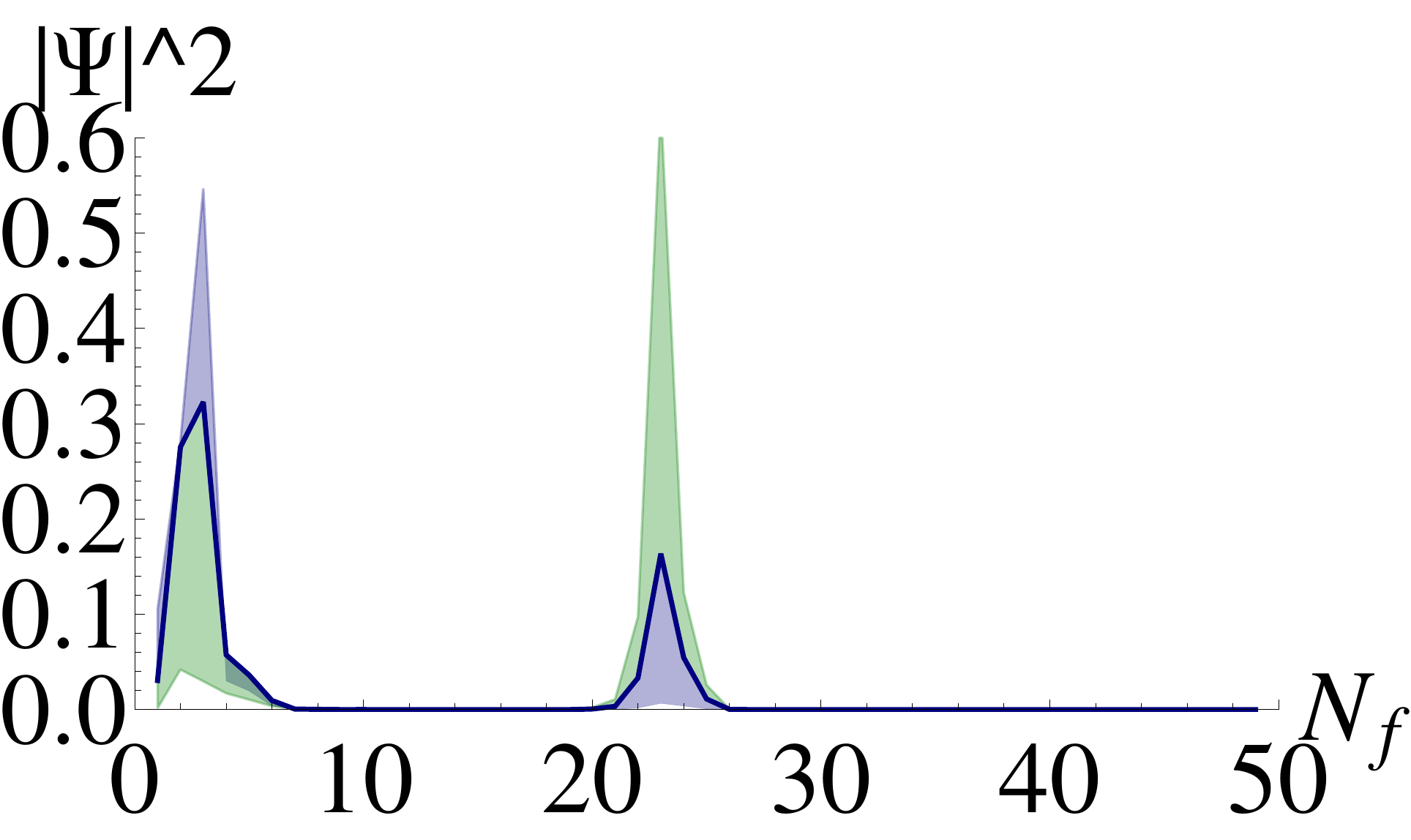}}
\subfigure[$\beta=0.10,\epsilon=0.5$]{\includegraphics[width=0.3\textwidth]{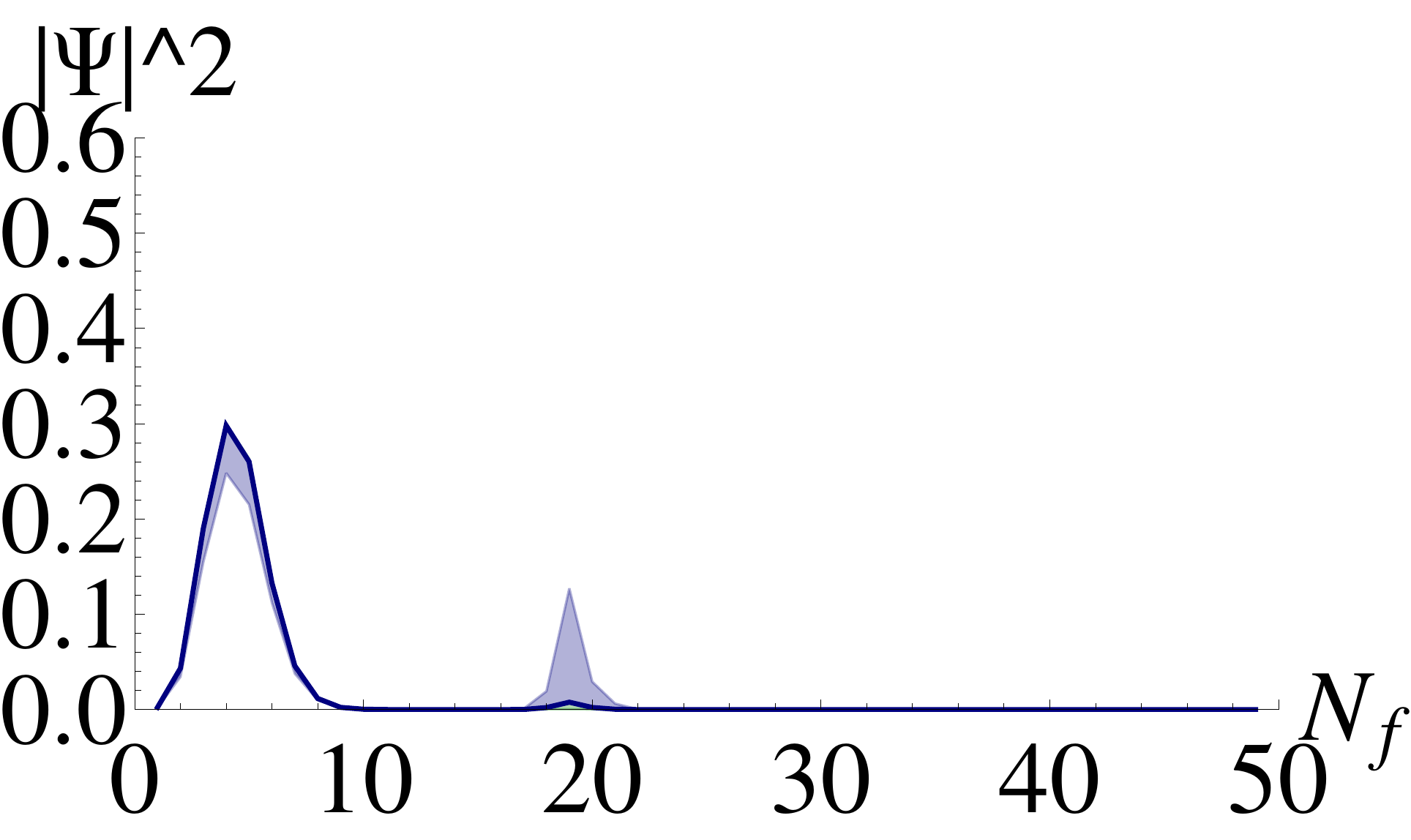}}
\subfigure[$\beta=0.11,\epsilon=0.5$]{\includegraphics[width=0.3\textwidth]{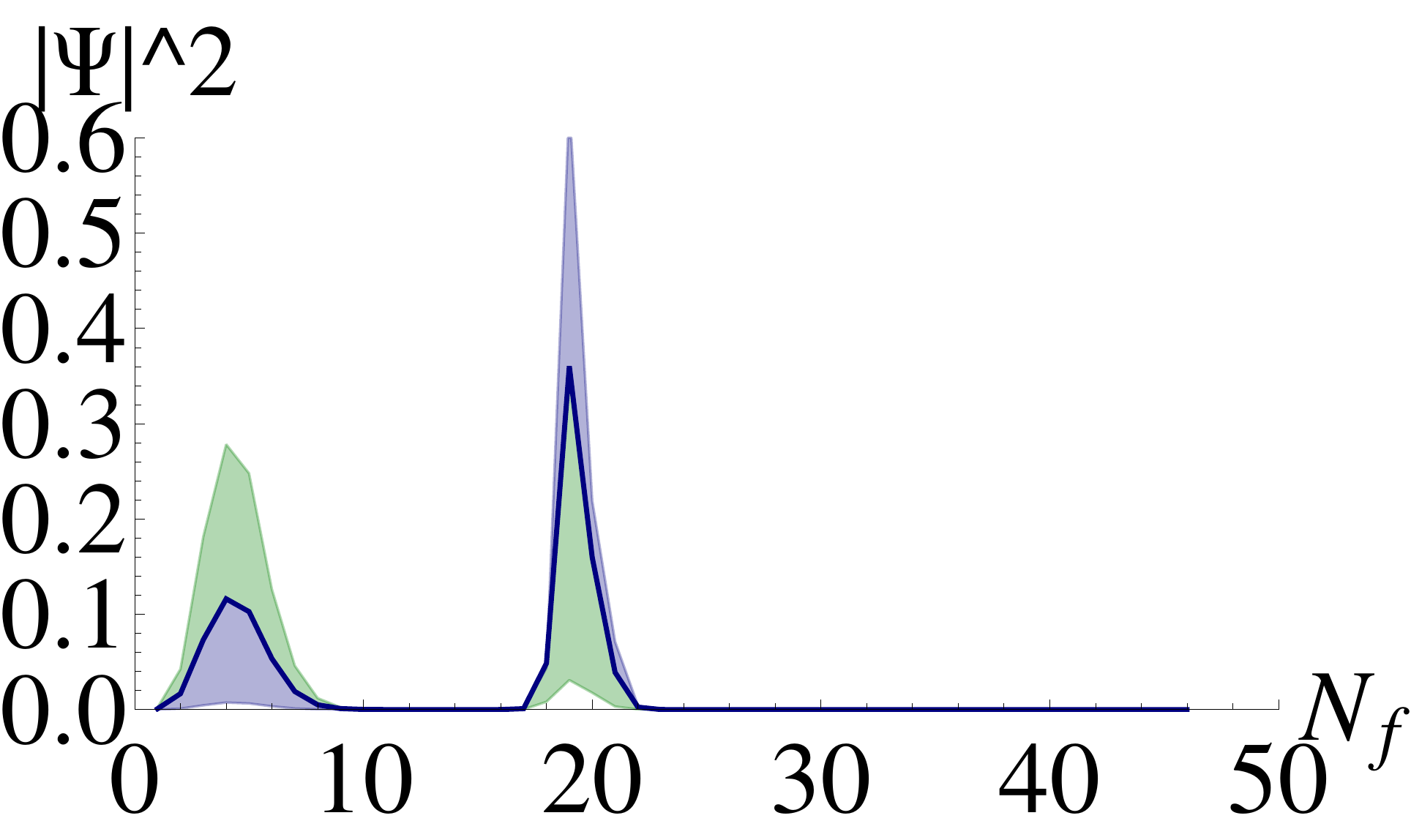}}
\subfigure[$\beta=0.12,\epsilon=0.5$]{\includegraphics[width=0.3\textwidth]{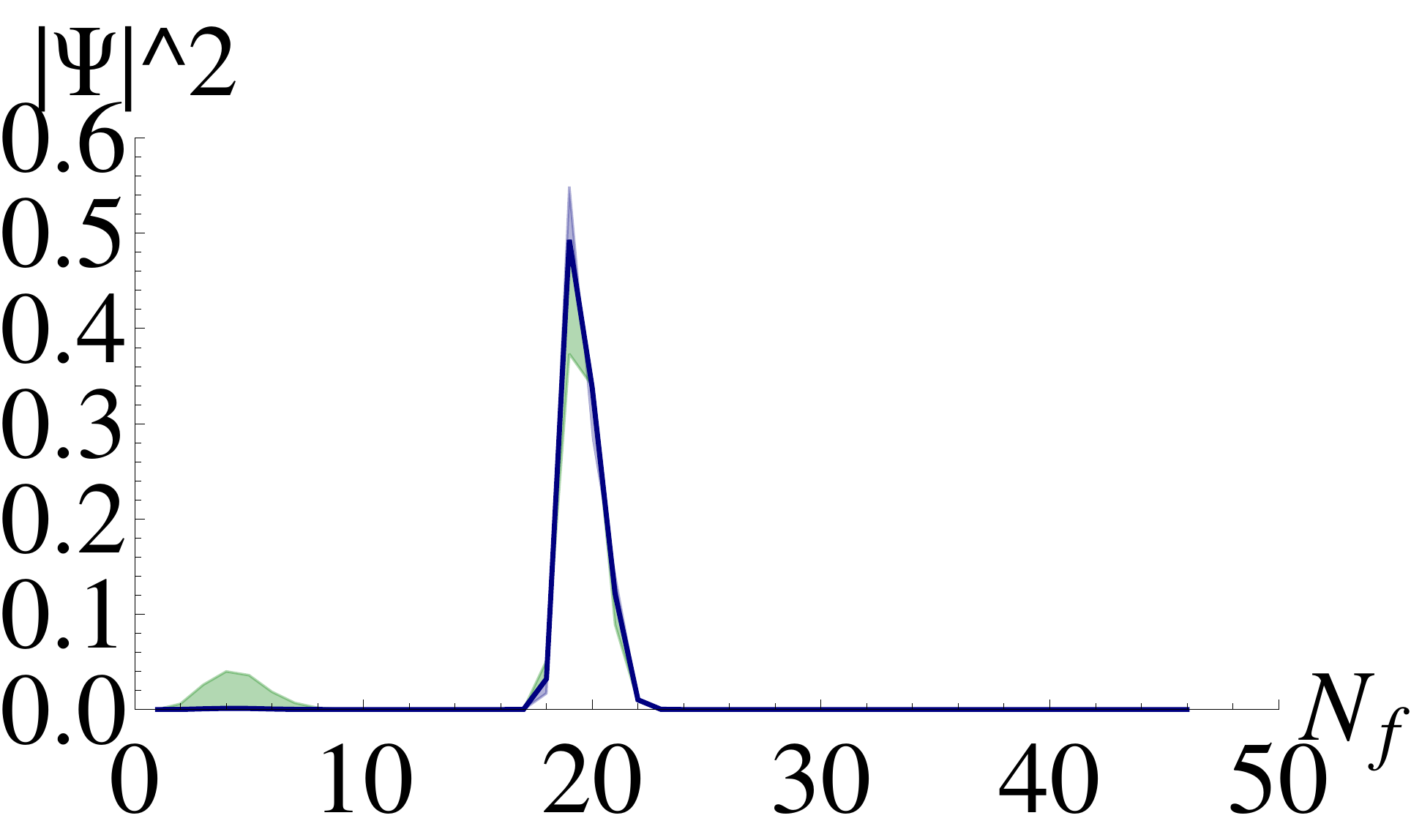}}

\caption{\label{fig:errorbars} The error in  $|\Psi_0(\final)|^2(\beta)$  for $\epsilon=0.12,0.5$.}

\subfigure[$\beta=7.5,\epsilon=0.12$]{\includegraphics[width=0.49\textwidth]{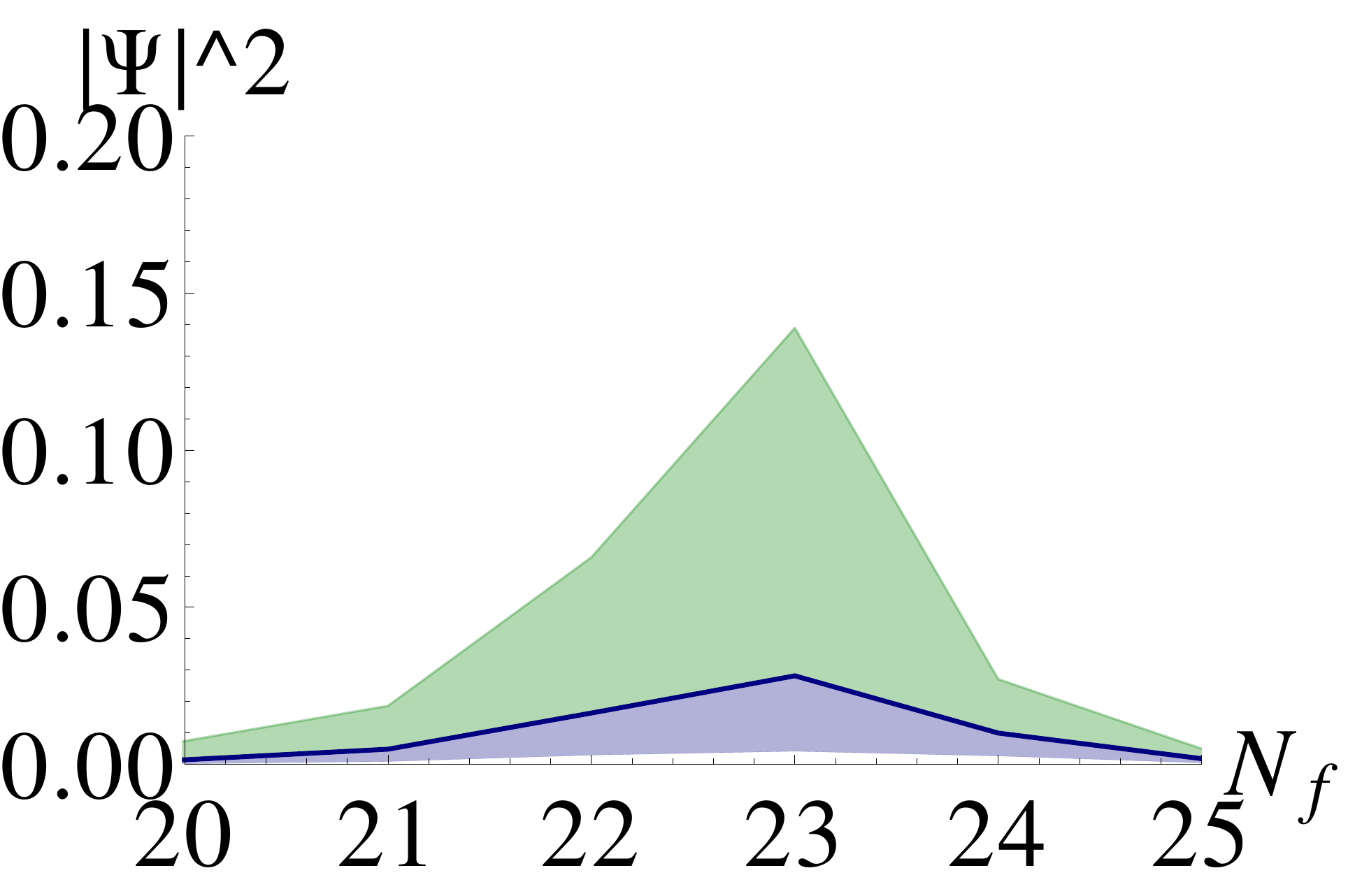}}
\subfigure[$\beta=8.5,\epsilon=0.12$]{\includegraphics[width=0.49\textwidth]{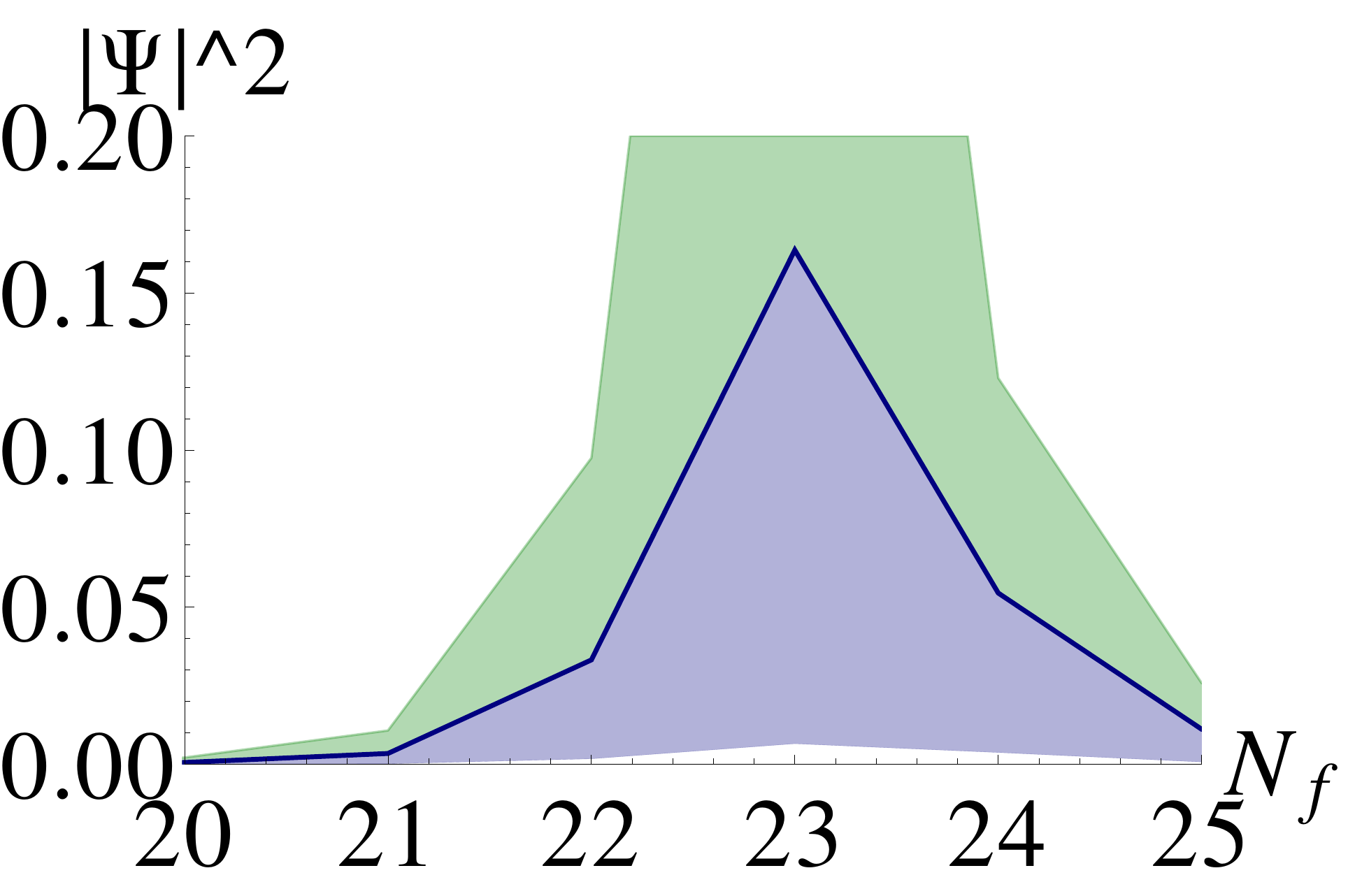}}
\caption{\label{fig:errorbars2} Close up of the peak at high $\beta$ to show that the peak also
  forms for the lower limit in the error.}
\end{figure}

The error in $|\Psi_0(\final)|^2$ is estimated from the errors in the interpolating functions
for $\avS_\beta(\final)$ and $\partn_0(\final)$ as shown in \mbox{Figure \ref{fig:errorbars}} for
\mbox{$\epsilon=0.12,0.5$}. The shaded region is the confidence interval for a confidence level of
$95 \%$  in our approximating function for $\avS_\beta(\final)$.  The green region is the difference
between the lower limit of the error in $\avS_\beta(\final)$ and the mean while the blue region is the
difference between the upper limit  in this error 
and the mean.  
 Thus, there is a  growth of the error around the phase transition since the lower limit  
 begins the phase
transition earlier than the upper limit.  For $\epsilon=0.12$ the appearance of the second peak in lower limit, green in \mbox{Figure
  \ref{fig:errorbars}}, coincides with the thermalisation limit.  In this case it does look as
though the second peak is dominated by the error. On the other hand, it is important to note that
the peak does start to develop in the lower limit as well.  We show this in Figure
\ref{fig:errorbars2}, where we have zoomed in to the peak of $|\Psi_0|^2$ for $\beta=7.5$ and $\beta=8.5$
.  Here, even the lower limit  slowly forms a peak.  For $\epsilon=0.5$ the thermalisation
limit occurs at a larger $\beta$ compared to the appearance of the second peak and the errors become small
enough post the phase transition, to see the dominance of the second peak. Similar analysis for the
confidence region for  $\partn_0(\final)$ shows the error to be subleading compared to the
uncertainty in the approximation of the average action.

\section{Results} 
\label{four}

The two different peaks in $|\Psi_0(\final)|^2$ correspond to two distinct discrete
geometries as we will see below.  Not surprisingly, these geometries strongly resemble the two phases
exhibited in \cite{2dqg}. The first peak at smaller $\beta$ corresponds to a 
a continuum phase and the second peak at larger $\beta$ corresponds to a non-continuum phase with
a  distinctive layered structure characteristic of the  crystalline phase of \cite{2dqg}.

\begin{figure}
\subfigure[\label{fig:randomone}$\final=4$, $\beta=0.2$]{\includegraphics[angle=45,width=0.5\textwidth]{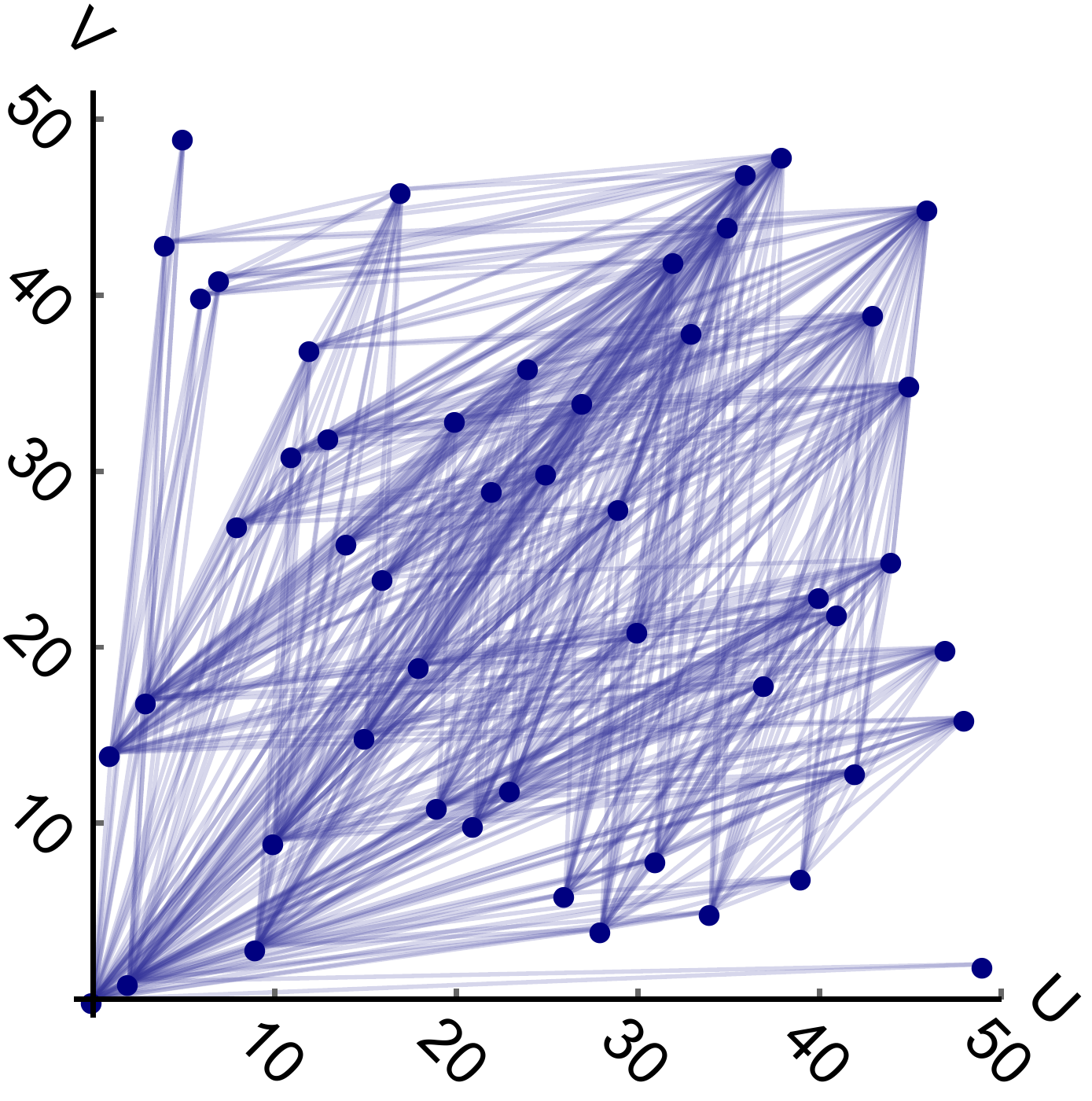}\hspace{10pt}}
\subfigure[\label{fig:Finalconfig2}$\final=23$, $\beta=7.6$]{\includegraphics[angle=45,width=0.5\textwidth]{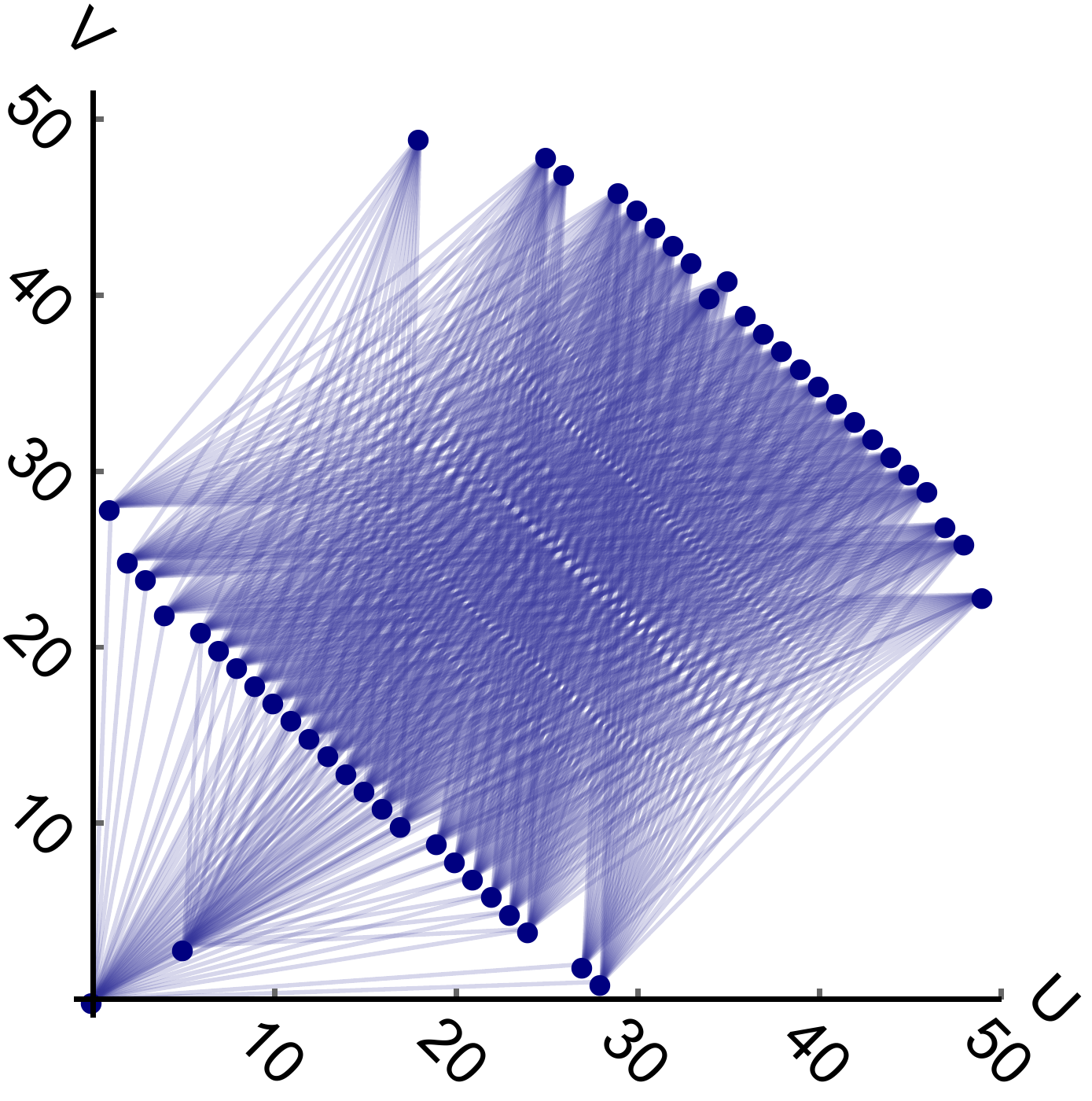}\hspace{10pt}}
\caption{Typical 2d orders at   $\epsilon=0.12$ associated with (a) the first and (b) the second
  peak,  in lightcone coordinates. The lines indicate causal relations between elements.} 
\label{fig:Finalconfig}
\end{figure}

\begin{figure}
\centering
\includegraphics[width=0.7\textwidth]{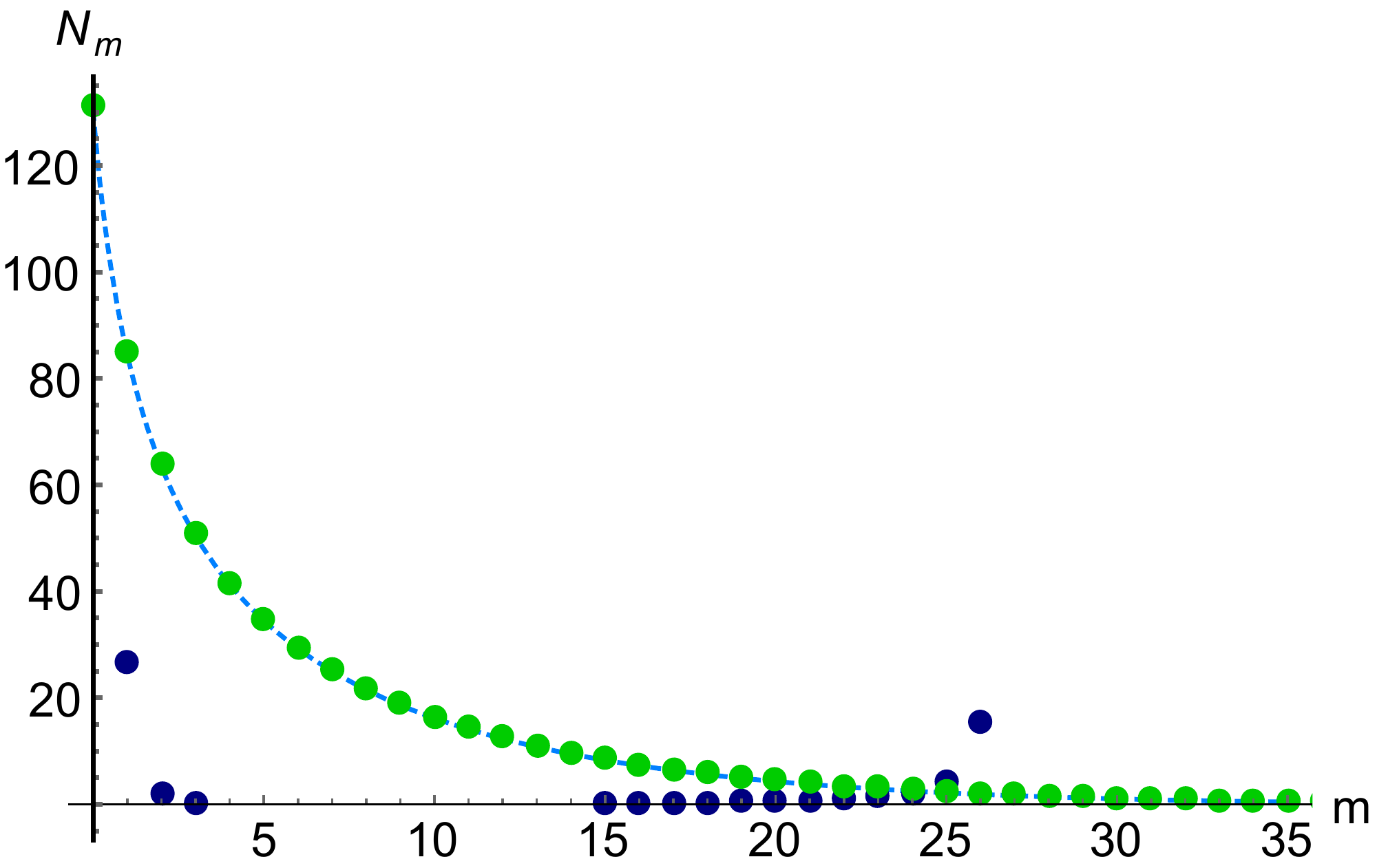}
\caption{\label{fig:Nirandom} The  green dots show $N_i$ for a causet in the first peak, for
  $\final=4,\beta=0.2$ and the blue dots for a causet from the second
  peak $\final=23,\beta=7.6$. The dotted line is the analytic calculation for sprinklings into  2d
  Minkowski spacetime.  Note: $N_0$  for the second peak geometry is $\sim 600$ and is not  visible in this plot.} 
\end{figure}

After locating the first and second peak values of $\beta$ and $\final$, we performed more extensive
MCMC simulations for a range of observables around these peaks for $\epsilon=0.12$.  They include
the proper-time or height of the 2d order, the distribution of the $N_i$ and the ordering fraction
(the ratio of the number of relations to the number of possible relations $\binom{N}{2}$.) The
causets in the first peak  around $\final \sim 4$ are,  predictably, those which are approximated by
2d Minkowski spacetime, i.e., they are random 2d orders. Figure \ref{fig:randomone} is an example of
a typical causet in this peak. In particular, the distribution of the $N_i$ is the
same as that of 2-d flat spacetime \cite{abundances}, , and the ordering fraction gives a Myrheim-Myer
dimension of 2.  The distribution of the $N_i$ are shown by the green dots in Figure
\ref{fig:Nirandom} which clearly follow those obtained from analytic calculations.   

The causets in the second-peak with $\final \sim 23$ share many of the features of the crystalline phase
of \cite{2dqg}.  Figure \ref{fig:Finalconfig2} shows a 2d order generated at the end of the MCMC
trial for $N_f=23, \beta=7.6$. 
In particular, they are non-manifold like as seen in the distribution of the $N_i$ (the blue
dots) in Figure \ref{fig:Nirandom}. The length of the longest chain (height) in these causets is small with $h \sim
4 $, while the bulk elements preferentially arrange themselves into a large antichain of size $\sim
\final$. Since the free parameter $\beta$ has a ready interpretation as an inverse temperature as in
the Euclidean path integral, the dominance at large $\beta$ of the non-continuum causets thus could
be taken to mean that they represent the ground state of the theory.  Indeed, what is surprising  is that though
these causets have no continuum counterpart, they nevertheless possess properties that have a ready
physical interpretation of particular significance to the observable universe.

The ratio of $\final$ to the height of the poset at the second peak is $\sim 6$ which means that
there is a rapid expansion from a single initial element to a large final antichain. A quick look 
at a first peak causet shown in Figure \ref{fig:randomone} shows that this ratio is less than one for
causets in the first peak.  In Table  \ref{tab:Nfoverheight} the ratio of $\final$ to the height of the poset is shown
for $\final=22,23,24$ in the second peak. 
\begin{table}
 \caption{ A table of the $\final/\mathrm{Height}$ for 2d orders around the second-peak at $\final=23$ for
$\epsilon=0.12$. }
\label{tab:Nfoverheight}
\centering
\begin{tabular}{ l l l l}
\toprule
$\beta$ & $\final=22$ &$\final=23$ & $\final=24$ \\ \midrule
$6.8$	&$4.53	\pm 0.03$	  &$	5.09	\pm 0.03 $&$	5.71	\pm 0.03 $\\
$7.2$	&$4.58	\pm 0.02$&$	5.26	\pm 0.03$&$	5.78	\pm 0.03 $\\
$7.6$	&$4.71	\pm 0.02$&$5.93	\pm 0.03$&$	5.95	\pm0.04 $\\
\bottomrule
\end{tabular}
\end{table}
A look at Figure \ref{fig:Finalconfig2} shows this  explicitly: most of
the elements in $\anti_f$ are just $3$ time steps away from the initial element. 
Thus,  despite being spatially large ($\final \sim N/2$), the universe is still very young.
\begin{figure}
\centering{\includegraphics[width=0.7\columnwidth]{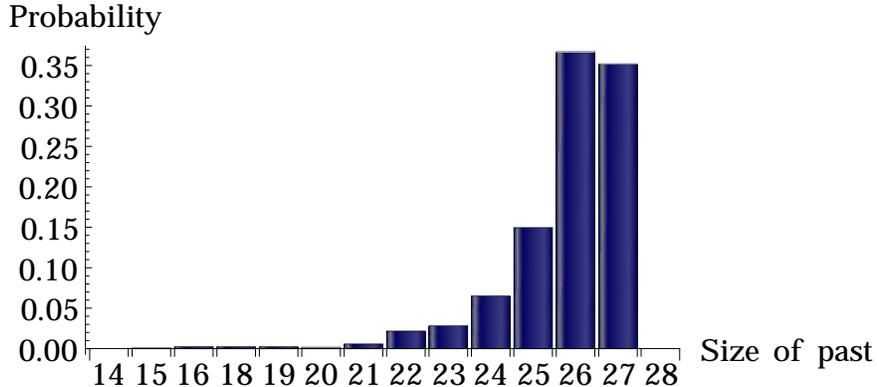}\hspace{10pt}}
\caption{The average distribution of past volumes of elements in $\anti_f$, for $\final=23$, $\beta=7.6$ and  $\epsilon=0.12$. } 
\label{fig:homogeneous}
\end{figure} 
Next,  Figure \ref{fig:homogeneous}  shows the probability distribution of the cardinality of the past of
the elements of $\anti_f$ at the second peak,  averaged over a sample of $50$ 2d orders in
the ensemble.   The distribution is  peaked around a past volume of $\sim 26$, falling off rapidly for smaller
volumes. Since the number of elements in the bulk of such causets is $26$ (excluding the initial
element) this means that there is a high degree of overlap in the pasts of each of the elements in
$\anti_f$ or high graph connectivity, given the constraints on $\infin$. This is clearly
illustrated  in Figure \ref{fig:Finalconfig}.

Taking these features together we find that an initial behaviour of the universe  which has much in
common with our expectations of the nature of the initial conditions coming  from the observable
universe.  This is particularly striking since the causets in the second peak are non-continuum like 
and have no continuum counterpart.  Each causet exhibits extensive past causal
contact between the elements of the final antichain, which is in stark contrast with the causal
structure of the standard FRW universe. This provides a discrete alternative to continuum
inflationary scenarios.    While the restriction to 2d is clearly unphysical, as in other approaches one
hopes to learn general lessons from it.  Thus 2d CST  explicitly demonstrates that the continuum may
be  inadequate to describe deep quantum gravity effects which could  nevertheless play a crucial role in
observable aspects of the early universe.  

It is useful to try to compare these results with the HH wavefunction in (a) Euclidean and (b)
Causal Dynamical Triangulations 2d quantum gravity \cite{david,lollambjorn}, (the latter
incorporates causality, although is not fundamentally discrete)  where the size of the final
hypersurface is represented by the length $L$ of the boundary circle.  In (a) the wavefunction has a
singularity at $L=0$ but dies out exponentially with increasing $L$. The singularity can be
attributed to the proliferation of baby universes when the cut-off is taken to zero
\cite{lollambjorn}.  In (b) while the singularity is tamed the contribution nevertheless peaks at
$L=0$ with a similar large $L$ behaviour.  This is in contrast with our results. 

We conclude this section with some open questions and future directions. 

The role played by $\beta$ in our analysis though non-trivial requires understanding.  In full 2d
quantum gravity $\beta$ is taken to be a Wick rotation parameter with $\beta \rightarrow -i \beta $
being the physically relevant quantum regime. Calculations with the Euclidean measure are assumed to
analytically continue to this quantum regime in a manner similar to quantum field theory.  In
contrast, since the HH wavefunction is {\it defined} as a Euclidean path integral there is no
essential need for a $\beta $ different from $1$ -- all it provides is an overall scaling of the
action.  While in higher dimensions $\beta$ can be absorbed into rescaling of $l_p$ this is not the
case in 2d quantum gravity because of the absence of a fundamental scale.  However, our analysis
clearly demonstrates that $\beta$ plays a physical role -- tuning $\beta$ shifts the peak
contributions from manifold like causets to non-manifold like causets. Recent work on the scaling
properties of 2d CST shows that both $\beta$ and $\epsilon$ play a significant role in the large $N$
behaviour of the theory, and it is plausible that a better understanding of $\beta$ lies in this
direction \cite{2drg}. A similar analysis for the HH-wavefunction by adding a parameter $\final$ to
the analysis (which would require much more extensive computational resources) could change the RG
flows of \cite{2drg} non-trivially, and lead to different  fixed points for $\beta$.  This is a
direction we hope to pursue in the near future.

Finally, the boundary term plays a crucial role in continuum formulations of quantum gravity.
It would be interesting to see how our results are affected by the recent proposal for a discrete
Gibbons Hawking term \cite{bdjs}.

\section{Discussion} 
\label{five}
\newcommand{\vol}{\mathrm{vol}} We now return to the question of  whether  $\Psi_0(\final)$
can be given  a covariant interpretation in the sum-over-histories framework.  We find that such an
interpretation is indeed possible in the quantum measure formulation \cite{qmeasure}.

\begin{figure}
\centering{\subfigure[]
{\includegraphics[width=0.20\textwidth]{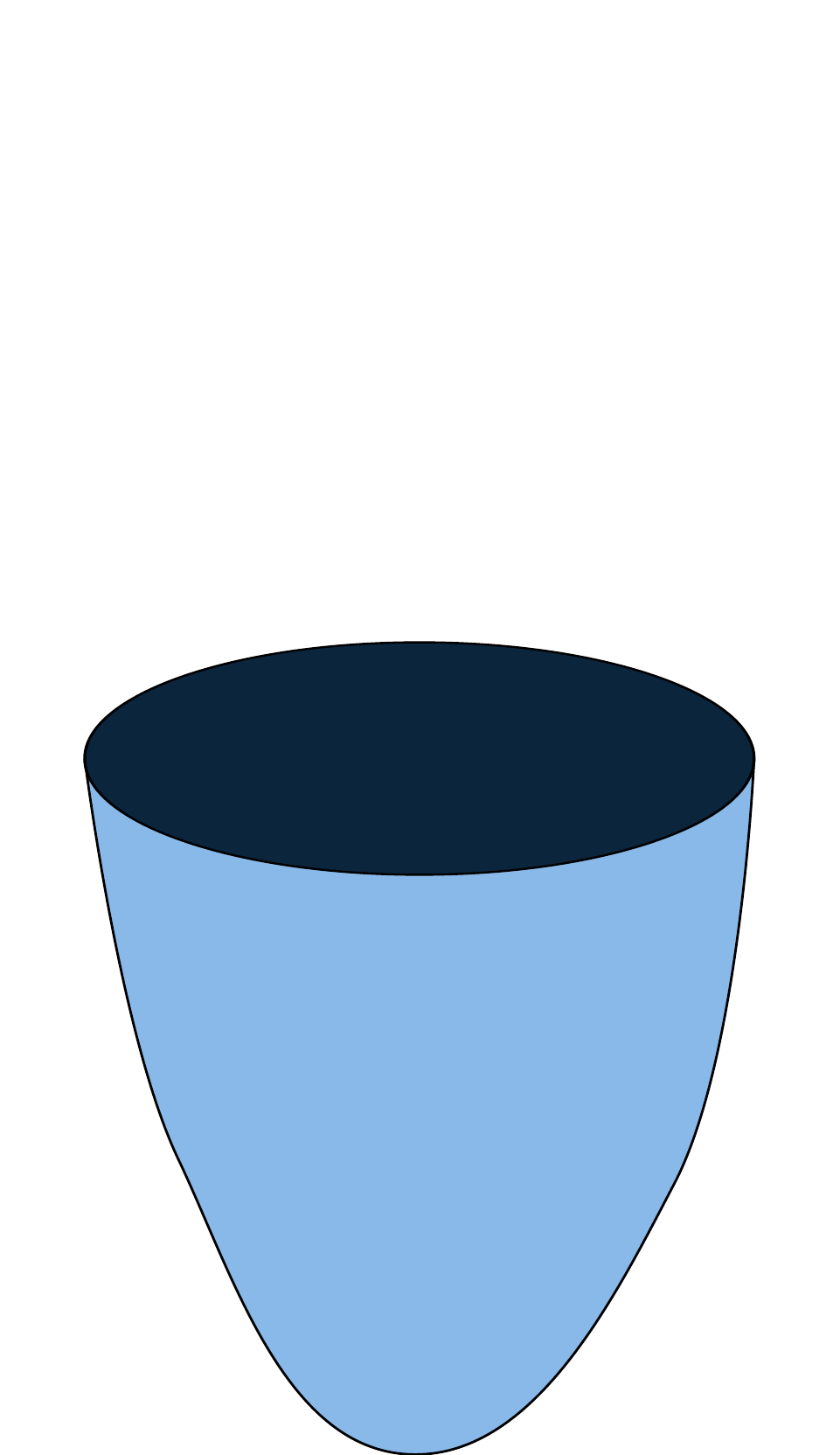}} 
\hskip0.1cm \subfigure[]{\includegraphics[width=0.2\textwidth]{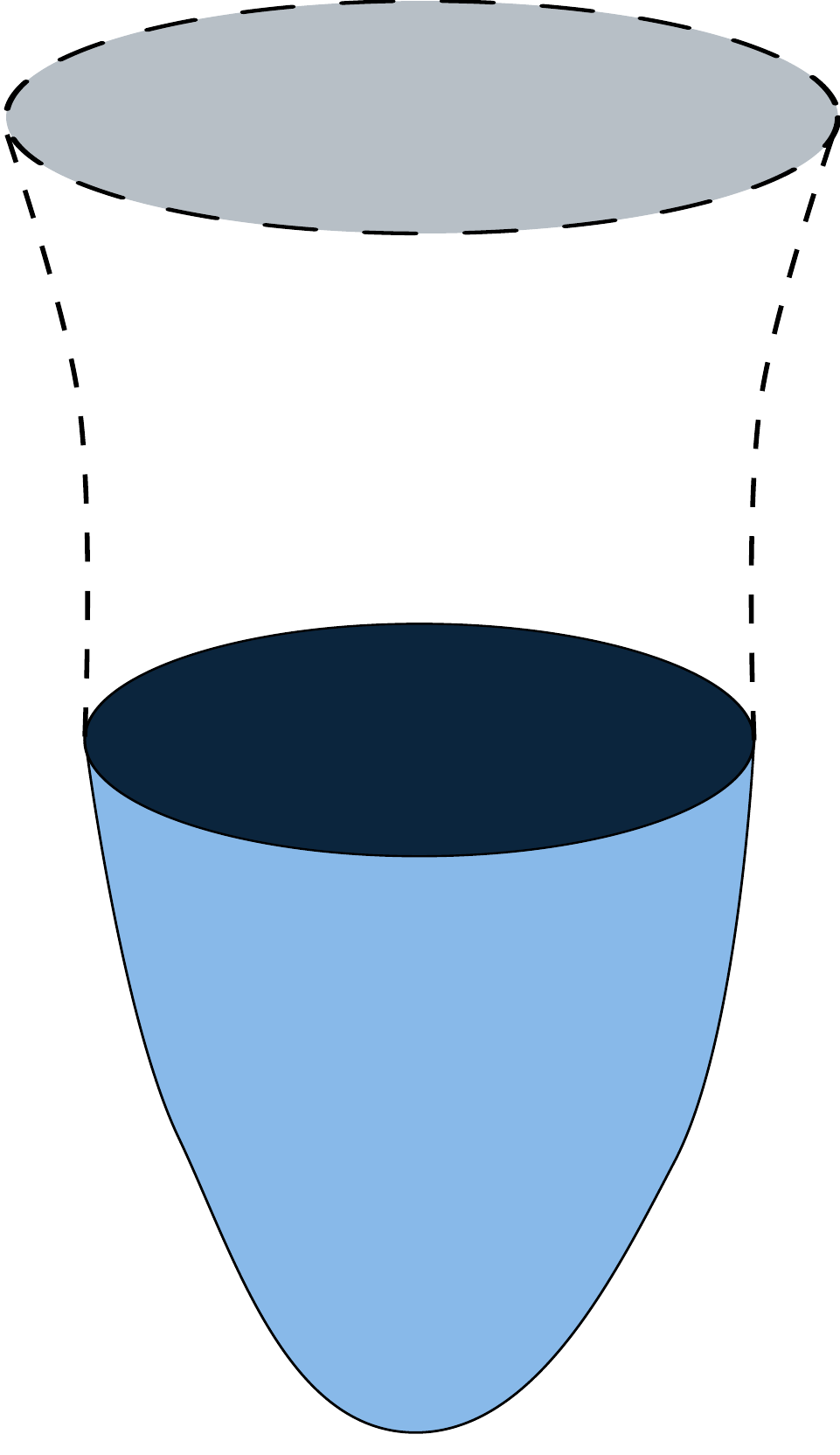}}
\subfigure[]{\includegraphics[width=0.35\textwidth]{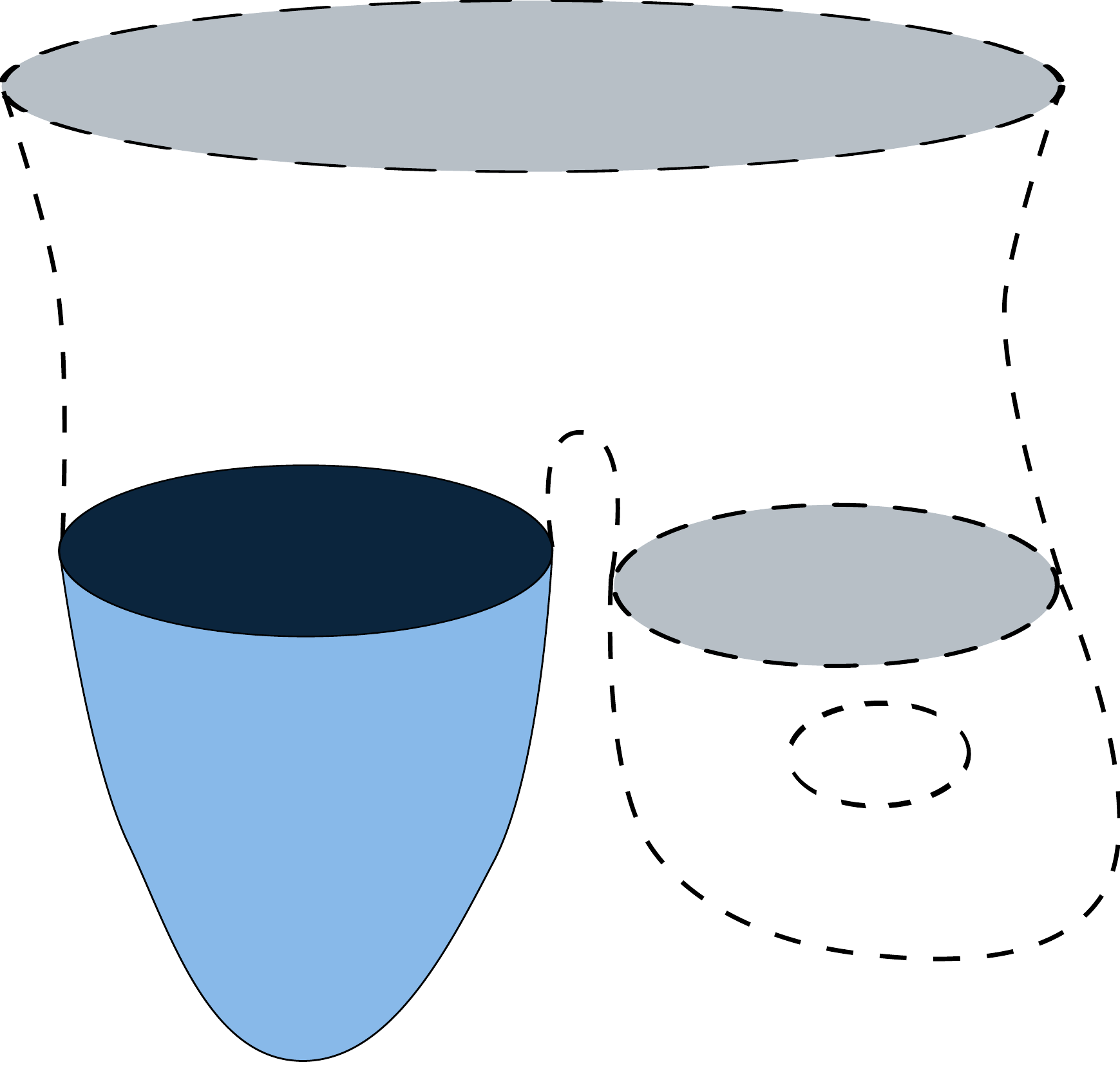}}}
\caption{\label{fig:hhfigs} Two possible subsequent evolutions (b) and (c) of an initial no-boundary
  spacetime (a).}
\end{figure}

We use the more familiar (but less concrete) language of the continuum to illustrate our proposal.
In the continuum the HH proposal is supposed to give the amplitude for a spatial initial condition
$\Psi_0(\Sigma, h)$. If the subsequent evolution occurs via a putative Hamiltonian dynamics with
$(\Sigma, h)$ a Cauchy hypersurface\footnote{While $(\Sigma, h)$ is obviously not the full initial
  data, the evolution of $\Psi_0$ will depend on the details of the canonical quantisation.}, any
subsequent evolution is constrained to depending only on the initial data on $\Sigma$.  However, the
transition from a formulation based on the path integral with possible topology change to one that
is purely Hamiltonian is somewhat an ad hoc hybrid approach. Instead, it is better to focus purely
on the sum-over-histories framework. Here, the no-boundary condition requires only that the path
integral is over histories with no initial boundary i.e., those that are topologically closed to the
past.  In particular (unless ad-hoc restrictions on topology are imposed) it is possible that
initially disconnected regions of spacetime could merge, thus rendering insignificant the role of a
particular ``final'' boundary $(\Sigma,h)$. Figure \ref{fig:hhfigs} shows two possible evolutions,
(b) and (c), of a no-boundary spacetime (a).
This illustrates the fact that  further conditions on the future evolution of the histories need to be imposed if
$\Psi_0(\Sigma,h)$ is meant to give information about the ``initial'' conditions of the
universe. For this $(\Sigma,h)$ must capture the ``complete'' information of the past. 

One way of doing this is to interpret $\Psi_0(\Sigma, h)$ not as the amplitude of the set of no-boundary
spacetimes with final boundary $(\Sigma,h)$ but of {\it all} no-boundary spacetimes $(M,g)$
containing at least one  spatial hypersurface $(\Sigma,h)$ which separates $(M,g)$ into its past and
future. Namely, we require that $M=J^-(\Sigma)\cup J^+(\Sigma)$ (where $J^\pm(S)$ denotes the causal
future and past of a set $S$) and such that $J^-(\Sigma)\cap J^+(\Sigma)=\Sigma$.
Since this set of histories contains no reference to a ``time'' label it is covariant. In the
language of measure theory, this set of histories forms a {\sl covariant event} with amplitude (or
{\sl quantum
measure})  $\Psi_0(\Sigma,h)$ \cite{qmeasure}.  In addition, we may define a  unimodular time
$T=\vol(J^-(\Sigma))$ for any separating hypersurface $(\Sigma, h)$ in a no-boundary spacetime.  We
can then similarly  interpret  $\Psi_0(\Sigma,h,T)$ to be the amplitude of the 
set of histories containing a separating  spatial hypersurface $(\Sigma,h)$ such that
$\vol(J^-(\Sigma))=T$. This too is covariant since $T$ is unique defined.   

In this continuum discussion we have side-stepped at least two important sets of questions. The
first is what the set of no-boundary spacetimes is; all we have done so far is specify that they are
topologically closed to the past. Should they also have finite past volumes? What are the
completeness requirements on these spacetimes?  The second is how a Euclidean signature
spacetime can be transformed into one of Lorentzian signature\footnote{One proposal is to match the
  signatures  by requiring that the extrinsic curvature on $(\Sigma,h)$ is identically zero \cite{Kzero}
.}. Neither of these poses a problem for CST: indeed, it is natural to consider finite element causal
sets that are past finite, and moreover, since causets are intrinsically Lorentzian, our framework
requires no ``signature matching conditions''.

In CST, covariance is implemented via label invariance. In our analysis of 2d CST, we have chosen to
count all relabelings so that the measure depends not only on the action but also on the number of
relabelings of a given 2d order.  This is a {\it choice} of measure or partition function(driven in
part by naturalness) which does not however affect covariance, since the physical observables,
including the action are purely covariant. Thus the HH wave function for a fixed $N$ is indeed
covariant since it is independent of the  relabelings of the 2d orders.

However, just as in the continuum there is an issue if one is to interpret the measure as being
merely that of the finite $N$ element causet.  If the causet evolves to a larger element causet, how
should we interpret $\Psi_0(\final)$? In particular $\final$ may no longer be an inextendible
antichain in the larger causet and hence $\Psi_0(\final)$ cannot be thought of as representing an
initial condition. This is similar to the conundrum in the continuum case whose resolution we have
sketched out. Rather than think of $\Psi_0(\final)$ as the measure on the $N$-element originary
causet with a future most antichain $\final$, it will be interpreted as the measure on the set
$\Omega$ of all originary {\it countable} labelled causets for which $\anti_f$ is separating, i.e.,
$\forall C \in \Omega, \,\, C=\mathrm{Past}(\anti_f) \cup \mathrm{Fut}(\anti_f) $ and
$\mathrm{Past}(\anti_f) \cap \mathrm{Fut}(\anti_f) =\anti_f $.  Following \cite{csg, observables},
the first step is to embed the finite sample space $\Omega^N$ of originary $N$ element causal sets
in the space $\Omega$.  The analogue of our prescription in the continuum would be to find the set
of all causets $C \in \Omega$ for which $\anti_f$ is separating, i.e., $C=\mathrm{Past}(\anti_f)
\cup \mathrm{Fut}(\anti_f) $, $\mathrm{Past}(\anti_f) \cap \mathrm{Fut}(\anti_f) =\anti_f
$. However, since $\Omega$ is the set of labelled causets, the question is whether such a set
corresponds to a covariant observable.

Following \cite{observables} we pose this question in the language of measure theory, where one
begins with the  triple
$(\Omega, \mathfrak A, \mu)$.  Here the  {\sl event algebra} $\mathfrak A$ over $\Omega$ is a
collection of subsets of $\Omega$ closed under the finite set operations of union, complementation
and intersection and includes $\Omega$ and the empty set.   $\mu$ is the measure on $\mA$ and can be
either classical or quantum. A  classical measure satisfies the sum rule 
\begin{equation} 
\mu(\alpha \sqcup \beta)=\mu(\alpha) + \mu(\beta)
\end{equation} 
for disjoint events $\alpha,\beta$, while a quantum measure satisfies the sum rule 
\begin{equation} 
\mu(\alpha \sqcup
\beta\sqcup \gamma)=\mu(\alpha\sqcup
\beta)+\mu(\alpha\sqcup\gamma) +\mu(\beta\sqcup \gamma) - \mu(\alpha) - \mu(\beta) - \mu(\gamma). 
\end{equation} 
for disjoint events $\alpha,\beta,\gamma$ \cite{qmeasure}\footnote{The quantum measure can also be cast as
a vector measure which satisfies the classical sum-rule.}. 
The set of observables is thus simply an element of the event algebra. However, since 
$\Omega$ is the set of  labelled causets, not all  choices of an event algebra will yield
covariant  observables. Although a non-covariant  event
algebra can be quotiented to form a label independent or  covariant algebra, the measure too should
be chosen to be label invariant. 
 
Following \cite{observables} we instead consider the covariant  event sigma\footnote{A sigma algebra
  is an event algebra which is in addition closed under countable set operations.} algebra $\Stem$ over
the set of unlabeled causets $\tO$ constructed as follows. A {\sl stem} $\sigma \subset C \in
\tO$ is a past-set i.e., $\mathrm{Past}(\sigma)=\sigma$ and a {\sl stem event} $\alpha_\sigma = \{ C
\in \tO| \sigma \,\, \mathrm{\, is \,\, a \,\, stem \,\, in\,\, }C \}$. A stem event is thus the set
of causets that possesses a particular past set or stem. The set of the stem events generate the
stem sigma algebra $\Stem$. Thus  a stem event is also an event in the set of labelled causal
sets $\Omega$ but is invariant under relabellings.

Define the set $\alpha^N_{HH}(\anti_f)\subset \tO$ as the set of causets in $\tO$ containing an
inextendable antichain $\anti_f$, with $|\mathrm{Past}(\anti_f)|=N-\final$.  $\anti_f$ is {\sl
  separating}, i.e., every element in such a causet lies either in $\anti_f $,
$\mathrm{Past}(\anti_f)\backslash \anti_f$ or $\mathrm{Fut}(\anti_f) \backslash \anti_f$. This is a
covariant characterisation, and indeed, we now show that $\alpha^N_{HH}(\anti_f)$ belongs to $\Stem$
and is therefore a covariant {\sl Hartle-Hawking} (HH) event.  We use arguments similar to those
in the Proposition in \cite{antichain}.

Begin with a finite causet $c_N\in \Omega_N$ with $\anti_f \subset c_N$ the complete set of its 
future most elements.  Let $d_M$, $M>N$ be a causet in $\Omega_M$ such that $c_N$ itself is a
stem in $d_M$, but $\anti_f \subset c_N$ is {\it not} inextendable in $d_M$ -- in other words,
$\anti_f$ does not divide $d_M$. It is clear that the event $stem(d_M)$ contains causets for
which $\anti_f$ is not inextendable, $\anti_f$ being a labelled set.   Further, if we
require that $|\mathrm{Past}(\anti_f)|=N-\final$, then $stem(d_M)$ contains no inextendable antichain 
$\anti_f$ of cardinality $\final$  with $|\mathrm{Past}(\anti_f)|=N-\final$.   Let $D_{c_N}$ denote the set of all such $d_M$
associated with $c_N$ for arbitrary $M>N$, and define $Q_{c_N}= \bigcup_{d \in D_{c_N}} stem(d) \in
\Stem$.  While, $Q_{c_N} \subset stem(c_N)$, it is the complement of this set
which we are interested in, since these causets  contain $\anti_f$ as an inextendable antichain with
$|\mathrm{Past}(\anti_f)|=N-\final$. This set,  $\Phi_{c_N}=stem(c_N)\backslash Q_{c_N}$ is also an element
of $\Stem$.  

Finally, consider the set $S_N$ of all possible $c_N \subset \Omega_N$ with $\anti_f$  an
inextendable antichain and $|\mathrm{Past}(\anti_f)|=N-\final$.  The HH event is then the set
$\alpha^N_{HH}(\anti_f) = \bigcup_{c_N \in S_N} \Phi_{c_N} \in \Stem$.  It is clear from this
construction that $\anti_f$ is indeed an inextendable, dividing antichain in any infinite element
causet in this set, with $|\mathrm{Past}(\anti_f)|=N-\final$, thus ensuring that no disjoint universe can
``join up'' at a coordinate time greater than $N$.  

\newcommand{\re}{\mathbb R} Thus the HH event $\alpha^N_{HH}(\anti_f)$ is indeed a  covariant event or
observable, and we may interpret our calculation of $\Psi_0(\final)$ (Eqn (\ref{hh})) as a
prescription for giving the measure on {\it this} class of observables.  As we have constructed it,
$\Psi_0(\final) \in \re$. We have moreover given no further information re. the nature of the
measure, i.e., whether it is quantum or classical.  Thus, specifying the measure on the
set of all HH-events will not suffice to give us the measure of other covariant events in $\Stem$
some of which could also be of physical interest.  Nevertheless, providing an covariant interpretation for the HH
wavefunction  seems a satisfying start to answering what is a very challenging set of
questions in quantum gravity, namely how to determine a fully {\it covariant} initial state of the
universe.

\vskip 1cm

MCMC simulations were conducted on the HPC cluster at the Raman Research Institute.
This work was supported in part  under an agreement with Theiss Research and funded by a grant from the
Foundational Questions Institute (FQXI) Fund, a donor advised fund of the Silicon Valley Community
Foundation on the basis of proposal FQXi-RFP3-1346 to the Foundational Questions Institute. 
This work was also supported by  funding from the European Research Council under
the European Union’s Seventh Framework Programme (FP7/2007-2013) / ERC Grant Agreement n.306425
``Challenging General Relativity''. 
LG was also supported by the ERC-Advance grant 291092, ``Exploring the Quantum Universe'' (EQU).

\bibliographystyle{utphys}

\bibliography{bibliography}

\end{document}